\documentclass[showpacs,aps,prd,longbibliography,nofootinbib,showkeys,superscriptaddress,twocolumn,longbibliography]{revtex4-2}
\usepackage{graphicx}
\usepackage{bm}
\usepackage{multirow}
\usepackage{natbib}
\usepackage{array}
\usepackage{amsmath}
\usepackage{mathrsfs}
\usepackage{amssymb}
\usepackage{float}
\usepackage{xcolor}
\usepackage{enumerate}
\usepackage{color}
\usepackage{slashed}
\usepackage{hyperref}
\usepackage{url}
\usepackage{txfonts}
\usepackage{mathtools}
\usepackage{booktabs}
\usepackage{BOONDOX-cal}
\usepackage{orcidlink}
\usepackage{longtable}
\usepackage{makecell}
\usepackage{cancel}

\allowdisplaybreaks[4]
\hypersetup{hypertex=true,colorlinks=true,linkcolor=blue,anchorcolor=red,citecolor=blue}

\newcommand{\dif}{\mathrm{d}}

\newcommand{\tds}{\tilde{s}}
\newcommand{\tdt}{\tilde{t}}
\newcommand{\tdu}{\tilde{u}}

\begin{document}
\title{Transport coefficients of dense nucleon matter at low temperature}
\author{Jianing Li\orcidlink{0000-0001-7193-7237}}\email{jianingli@ccnu.edu.cn}
\affiliation{Key Laboratory of Quark and Lepton Physics (MOE) and Institute of Particle Physics, Central China Normal University\\ Wuhan 430079, China}
\author{Weiyao Ke\,\orcidlink{0000-0002-5630-2388}}\email{weiyaoke@ccnu.edu.cn}
\affiliation{Key Laboratory of Quark and Lepton Physics (MOE) and Institute of Particle Physics, Central China Normal University\\ Wuhan 430079, China}
\date{\today}
\begin{abstract}
	The transport property of cold and dense nucleon matter is important for nuclear physics but is relatively less studied than that at finite temperatures. In this paper, we present a primary study of bulk and shear viscosities in the limit $T/\mu_B \ll 1$, where $T$ and $\mu_B$ are the temperature and the baryon chemical potential. The analysis is performed for a generic system where nucleons are dressed by the condensation of both scalar and vector interactions. Under the relaxation time approximation of the Boltzmann equation, we compute the viscosities of the system to leading power in $T/\mu_B$ expansion and establish a relation between the thermodynamic potential and transport coefficients, including bulk viscosity ($\zeta$) and shear viscosity ($\eta$). It is found that hydrodynamic stability ($\zeta>0$) imposes additional constraints on the thermodynamic potential. As an example, these relations are applied to the Walecka model. The fluid properties of the cold and dense nucleon matter are characterized by the dimensionless combination of viscosities times the quasi-Fermi momentum over the enthalpy. Furthermore, we discuss the implication of the stability condition on the range of applicability of the model.
\end{abstract}
\maketitle
\section{Introduction}
The exploration of cold dense nucleon matter is crucial for advancing our understanding of nuclear dynamics under conditions near nuclear saturation density and at relatively low temperatures. This regime is particularly relevant for the upcoming Electron-Ion Collider (EIC) in the United States \cite{Accardi:2012qut} and the Electron-Ion Collider in China (EicC) \cite{Anderle:2021wcy}, as well as for phenomena in low-energy nuclear reactions \cite{Ishikawa:2003ne,Pang:2012,Yuan:2012cj,Shi:2011rh}. Unlike relativistic heavy-ion collisions, which explore the properties of hot quark-gluon plasma (QGP) at high temperatures \cite{Andronic:2009gj,Muller:2012zq,Shuryak:2014zxa,ALICE:2016ccg,Busza:2018rrf,Bzdak:2019pkr}, EIC and EicC experiments leverage high-energy electron-nucleus collisions to probe the internal dynamics of quantum chromodynamics (QCD) in nucleons and nuclei \cite{Ethier:2020way,AbdulKhalek:2021gbh}. These facilities also offer a unique opportunity to study the transport properties of strongly interacting matter which play critical roles in the behavior of confined hadronic matter \cite{Shen:2022oyg,Zhao:2022ayk,Schenke:2024cnc}.

The QCD media in EIC and EicC experiments are cold nuclei, where phenomena such as color confinement, nucleonic degrees of freedom, nucleon correlations, and shell effects dominate \cite{Abir:2023fpo}. This is in stark contrast to the quark-gluon-dominated physics of the hot QGP produced in heavy-ion collisions \cite{Heinz:2004qz}. The QGP is a finite-temperature, ultrarelativistic system characterized by a thermal abundance of nearly massless gluons \cite{Pasechnik:2016wkt}, whereas nucleon matter at low temperatures is primarily a fermionic system. While the equation of state of cold nuclear matter has been extensively studied in various contexts \cite{Burgio:2020fom}, the transport properties of cold nucleon matter governed by strong interactions, as probed by low-energy nuclear reactions and electron-ion interactions, remain less well understood \cite{Deng:2023rfw}. 

Due to the formation of a Fermi surface in a cold fermionic system, the relevant low-energy degrees of freedom are confined to the vicinity of this surface in momentum space. This simplification allows the use of various theoretical tools to effectively reduce the description of the system from the full bulk momentum space to the dynamics on the Fermi surface \cite{Delacretaz:2022ocm}.
In this paper, we investigate the bulk viscosity ($\zeta$) and shear viscosity ($\eta$) of cold nuclear matter at low temperatures using the framework of Fermi liquid theory~\cite{Landau:1956yop}. We assume a temperature regime that is low enough to maintain a well-defined Fermi surface but not so low as to induce significant nucleon pairing.

Finally, while neutron stars also harbor cold, dense nuclear matter—with densities several times higher than the nuclear saturation density \cite{Tsang:2023vhh,Annala:2023cwx,Douchin:2001sv} and temperatures ranging from 1–10 MeV for newly born neutron stars \cite{van1991neutron,Yakovlev:2004iq,Page:2005fq} to 10–50 MeV for neutron star merger events \cite{Evans:2017mmy,LIGOScientific:2017ync}—the timescales of their dynamics differ drastically from those of nuclear reactions.
In nuclear collider experiments, the characteristic timescale of strong interactions is on the order of $10^{-20}$ seconds, and electroweak processes are negligible \cite{Paschos:2007pi}. In contrast, the interiors of neutron stars evolve over millisecond timescales, with weak interaction processes dominating phenomena such as cooling \cite{Schmitt:2017efp,Hoyos:2021uff,Hoyos:2020hmq,CruzRojas:2024etx,Harris:2024evy}, magnetic field evolution \cite{goldreich1992magnetic,Igoshev:2021ewx}, and gravitational wave emission \cite{Lindblom:1998wf,Lindblom:2001hd}.
Consequently, the strong-interaction viscosity considered in this paper, while relevant to the short timescales of nuclear reactions, does not play a significant role in the long-term evolution of neutron stars. However, the theoretical and computational methods developed here may be adapted in the future to study electroweak processes, which are critical to understanding neutron star dynamics \cite{Delacretaz:2025ifh}.

The rest of the paper is organized as follows. In Sec. \ref{sec1}, we derive expressions for these transport coefficients using the relaxation time approximation (RTA). Given the non-perturbative nature of QCD in this regime—and the limitations of lattice QCD due to the sign problem \cite{Aarts:2012yal}—we employ phenomenological models to explore this challenging domain. Specifically, we utilize the Walecka model to characterize the static properties of nucleonic matter \cite{Walecka:1974qa,Csernai:1986qf,Herbert:1992kd,DasGupta:2000md} in Sec. \ref{sec2}. The relaxation time is estimated by calculating tree-level scattering processes within the Walecka model framework. Finally, in Sec. \ref{sec3}, we present our conclusions and discuss potential future research directions.
\section{Transport coefficients of cold and dense fermions}\label{sec1}
\subsection{Relaxation time approximation with $T/\mu\ll 1$}\label{sec1a}
The relativistic Boltzmann equation for a single-component system in the presence of a collision term $\mathcal{C}\left[f\right]$ is written as~\cite{DeGroot:1980dk}
\begin{align}\label{eq:Boltzmann}
	\left(\frac{\partial}{\partial{t}}+\frac{\boldsymbol{p}}{E_{\boldsymbol{p}}}\cdot\nabla_{\boldsymbol{x}}-\nabla_{\boldsymbol{x}}E_{\boldsymbol{p}}\cdot\nabla_{\boldsymbol{p}}\right)f=\mathcal{C}\left[f\right]\,,
\end{align}
where we consider $f \equiv f(x^{\mu}, p^{\mu})$ is the fermionic distribution function, with $x^{\mu} = (t, \boldsymbol{x})$ and $p^{\mu} = (E_{\boldsymbol{p}}, \boldsymbol{p})$ representing the spacetime coordinates and four-momentum of the fermions, respectively. The zeroth component of $p^{\mu}$ is the energy $E_{\boldsymbol{p}} = \sqrt{\boldsymbol{p}^2 + m^2}$, where $m$ is the mass. With the metric defined as $g^{\mu\nu} = \mathrm{diag}(1, -1, -1, -1)$, the symmetric energy-momentum tensor is expressed as
\begin{align}\label{eq:EMT1}
	T^{\mu\nu}=-pg^{\mu\nu}+\left(p+\varepsilon\right)u^{\mu}u^{\nu}+{\delta}T^{\mu\nu}\,,
\end{align}
where $p$ is the pressure, $\varepsilon$ is the energy density, and $u^{\mu}$ is the four-velocity of the fluid cell. Furthermore, the dissipative term $\delta T^{\mu\nu}$ satisfies the Landau--Lifshitz condition $u_\mu \delta T^{\mu\nu} = 0$.

At the first order of the gradient expansion, we have
\begin{align}\label{eq:dEMT1}
	\Pi^{\mu\nu}=\eta\left(D^{\mu}u^{\nu}+D^{\nu}u^{\mu}+\frac{2}{3}\Delta^{\mu\nu}\vartheta\right)-\zeta\,\Delta^{\mu\nu}\vartheta\,,
\end{align}
where we define $\vartheta=\partial\cdot{u}$. Here, $\Delta^{\mu\nu} = u^\mu u^\nu - g^{\mu\nu}$ is the projection tensor, and $D_{\mu} = \partial_\mu - u_\mu u\cdot\partial$ is the derivative normal to $u^{\mu}$. It is useful to express these projections in the local rest frame, i.e., $u^\mu = (1, 0, 0, 0)$:
\begin{align}\label{projection}
	\Delta^{0\nu}=0\,,\quad
	\Delta^{ij}=\delta^{ij}\,,\quad
	D_0=0\,,\quad
	D_i=\partial_i\,.
\end{align}
The prefactors $\eta$ and $\zeta$ are the \emph{shear viscosity} and \emph{bulk viscosity}, respectively. 

In a system of well-defined quasiparticles, the energy-momentum tensor (neglecting residue interactions) is expressed as
\begin{align}\label{eq:EMT2}
	T^{\mu\nu}=\sum_{i}\int\frac{\dif^3\boldsymbol{p}}{\left(2\pi\right)^3}\frac{{p^\mu}{p^\nu}}{E_{\boldsymbol{p}}}f_i\,. 
\end{align}
where $i$ sums over all quantum numbers with degeneracies.
For a system of free relativistic fermion, the equilibrium distribution is typically given as\footnote{Due to the general formula presented in this section, the chemical potential is represented by $\mu$ instead of specifying a particular baryon type $\mu_B$ in Sec. \ref{sec2}.}
\begin{align}\label{eq:f1}
    \begin{split}
	f_0\left(x,p\right)&=\frac{1}{\exp\left[\beta\left(u\cdot{p}-\mu\right)\right]+1}\\
	&=\Theta(\mu-u\cdot p) + \mathcal{O}(T^2/\mu^2)\,,
    \end{split}
\end{align}
where $\beta = 1/T$ is the inverse of temperature $T$, and $\mu$ is the chemical potential of a conserved charge.\footnote{For antiparticles, $\mu$ is replaced by $-\mu$ in Eq.~(\ref{eq:EMT2}) However, for cold systems with $\mu/T\gg 1$, we will neglect the antiparticle contributions hereafter.}. In the second line, we have power expanded the distribution function in $T/\mu$.
Assuming that $f$ deviates slightly from equilibrium, we can decompose the particle distribution into $f = f_0 + \delta f$, where $\delta f$ is the nonequilibrium correction term.

\begin{figure}[htpb!]
	\centering
	\includegraphics[width=0.47\textwidth]{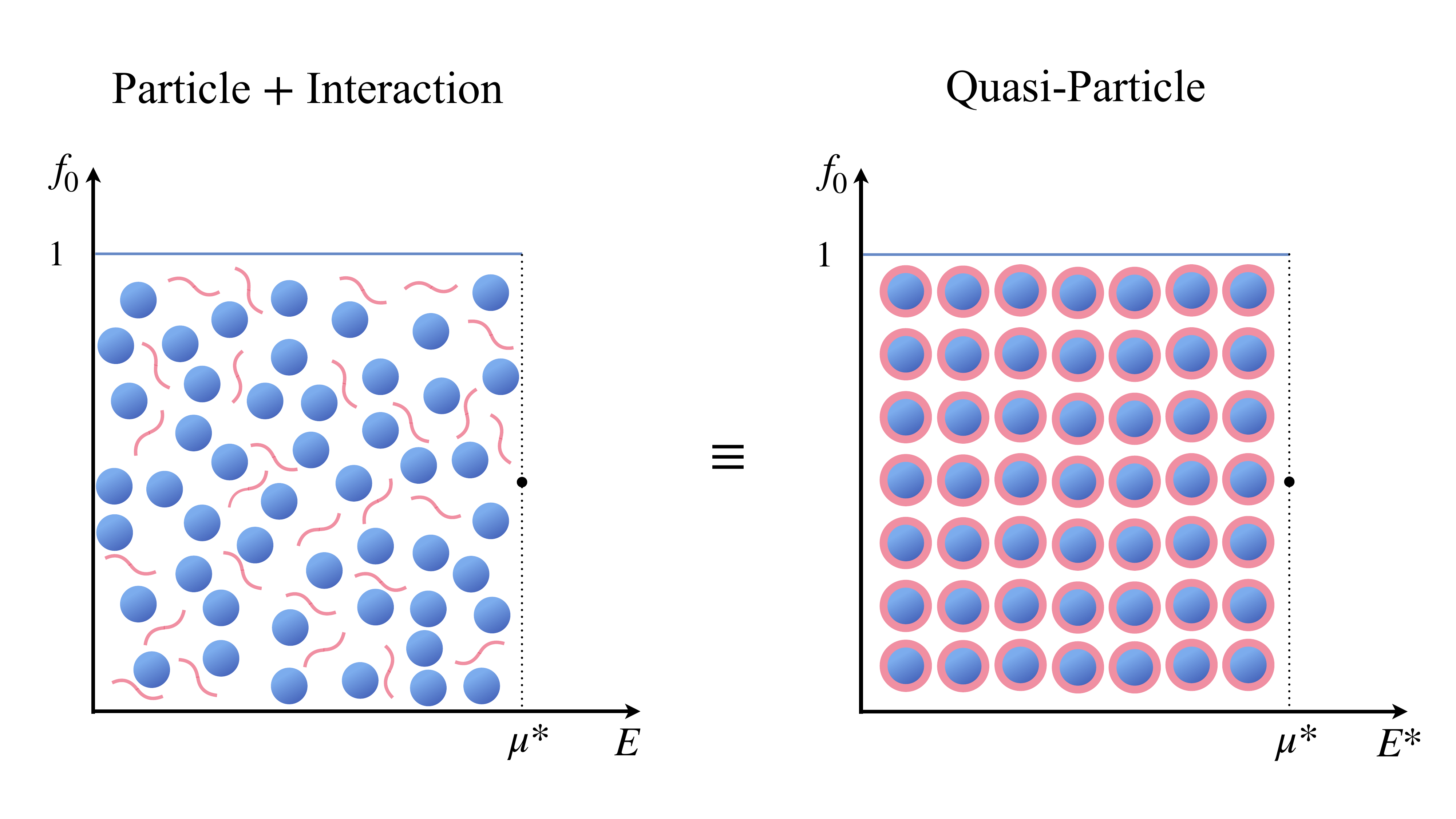}
	\caption{A schematic representation of Fermi liquid at zero temperature. Left: Bare particles and interactions. Right: Quasiparticles with effective mass $m^*$ and effective chemical potential $\mu^*$.
	\label{fig:quasi-particle}}
\end{figure}

Even though Eq.~(\ref{eq:f1}) describes a noninteracting many-particle system, which differs from a real-world system, it provides qualitative explanations for salient features of the latter. In 1956, Lev Landau introduced the \emph{Landau Fermi liquid theory} (LFLT) to explain the low-temperature behavior of interacting fermion systems, such as electrons in metals~\cite{Landau:1956yop}. The theory centers around the concept of quasiparticles, fermionlike excitations, that retain essential properties of free fermions but are modified by interactions~\cite{Pines1989:theory, Nozieres1998:theory}. The system’s behavior can still be described using Fermi--Dirac statistics with an effective mass $m^*$ and an effective Fermi surface $\mu^*$ modified by interaction, as illustrated in Fig. \ref{fig:quasi-particle}.
This theory has been instrumental in explaining various physical properties of metals and liquid helium-3, including specific heat and magnetic susceptibility, making it a cornerstone of condensed matter physics~\cite{Leggett:1975te}.
There are also numerous LFLT applications to study nuclear matter at the finite density from low-energy effective models of the strong interaction \cite{Baym:1975va,Klevansky:1992qe,Schwenk:2004hm,Epelbaum:2008ga,Baldo:2011gz,Friman:2012ft,Holt:2012yv,Holt:2013fwa,Ropke:2017pvg,Holt:2017uuq,Friman:2019ncm,Drischler:2021kxf}. 

Usually, in the mean-field approximation (MFA) of such an effective model for relativistic fermions, modifications in $m^*$ and $\mu^*$ are introduced due to spontaneous symmetry breaking in the scalar and vector channel interactions, respectively. Without loss of generality, we assume
\begin{align}
	\label{eq:dress function}
	\begin{gathered}
		m^*=m+\mathcal{f}\left(\bar\sigma\right)\,,
		\quad
		\mu^*=\mu+\mathcal{g}\left(\bar{\omega}_0\right)\,.
	\end{gathered}
\end{align}
Here, $\bar{\sigma}$ and $\bar{\omega}_0$ are the scalar and vector condensates that account for spontaneous symmetry breaking. For simplicity, the vector condensate has been rotated to the zeroth component. The functions $\mathcal{f}(\cdot)$ and $\mathcal{g}(\cdot)$ are dressing functions for the scalar and vector channels.

At zero temperature, the local equilibrium distribution function is given by $f_{0} = \Theta(\mu^* - u \cdot p)$, where $E^*_{\boldsymbol{p}} = \sqrt{{m^*}^2 + \boldsymbol{p}^2}$. The effective quantities $\mu^*$ and $m^*$ both depend on the chemical potential $\mu$. In a system out of the global equilibrium, we consider a chemical potential $\mu(x^\mu)$ that slowly varies in spacetime, thus
 both $\mu^*[\mu(x^\mu)]$ and $m^*[\mu(x^\mu)]$ are functions of the spacetime coordinates as well. 
Out-of-local-equilibrium effects can be thought of as the distortion of the shape of the Fermi surface at each point in spacetime. At finite but low temperatures, we can power expand the calculations in the small parameter $T/\mu^*$ and the arguments above still hold for at the zeroth order.

For long-wavelength excitations at zero temperature, Ref.~\cite{Delacretaz:2022ocm} shows that (for the nonrelativistic system) the deviation from the local equilibrium distribution $f_{0}$ could be parameterized as deformations of the Fermi surfaces,
\begin{align}\label{eq:coadjoint-orbit}
	\begin{split}
			&\quad f_0+\delta{f}(T=0)\\
			&=\Theta\left(p_F-|\boldsymbol{p}|\right)+ \delta\left(p_F-|\boldsymbol{p}|\right)\boldsymbol{n}_\theta \cdot \nabla_{\boldsymbol{x}}\Phi\left(\boldsymbol{x}, \boldsymbol{n}_\theta\right) + \cdots\,,
	\end{split}
\end{align}
where $\boldsymbol{n}_\theta=\boldsymbol{p}/|\boldsymbol{p}|$ is a unit vector and the structure of the deformation term follows from the canonical symmetry of the transport part in the Boltzmann equation. At vanishing temperature, nonforward scatterings are completely suppressed and the systems have an infinite number of conserved quantities and their corresponding hydrodynamic modes, labeled by $\theta$. At finite temperature, off-forward scatterings introduce small but finite breaking of the canonical transformation symmetry to the Boltzmann equation; therefore, one expects (1) a reduction of the number of conserved quantities, (2) the form in Eq.~(\ref{eq:coadjoint-orbit}) will be generalized by terms that were previously not allowed by canonical symmetry, and (3) such terms are suppressed by powers of $T/\mu$. For this reason, we start from a more general form of deviations around the Fermi surface,
\begin{align}\label{eq:finite-T-deltaf}
f_0+\delta{f}(T\neq 0) = \Theta\left(\mu^*-u\cdot{p}\right) + \delta\left(\mu^*-u\cdot{p}\right)\phi + \cdots
\end{align}
with a consideration of the Lorentz covariance.

Returning to the Boltzmann equation in Eq.~(\ref{eq:Boltzmann}), we first power expand the distribution function in $T/\mu$, and then write down the equation to the first order of the deviation $\delta f$,
\begin{align}
	\begin{split}
		-\frac{1}{E^*_{\boldsymbol{p}}}&\left[p^\mu \left(p\cdot\partial\right){u_\mu}-\frac{\dif{\mu^*}}{\dif{\mu}}p\cdot\partial{\mu}\right.\\
		&\,\left.+m^*\frac{\dif{m^*}}{\dif{\mu}}u\cdot\partial\mu\right]\delta\left(\mu^*-u\cdot{p}\right)=\mathcal{C}\left[\delta{f}\right]\,,
	\end{split}
\end{align}
Adopting the RTA, the collision term can be expressed as $\mathcal{C}\left[f\right]=-\delta{f}/\tau_\text{rel}$, where the relaxation time $\tau_\text{rel}$ depends on the detail of interaction. Then, $\phi$ is solved by
\begin{align}
	\phi=\frac{\tau_\text{rel}}{E^*_{\boldsymbol{p}}}\left[p^\mu \left(p\cdot\partial\right){u_\mu}-\frac{\dif{\mu^*}}{\dif{\mu}}p\cdot\partial\mu+m^*\frac{\dif{m^*}}{\dif{\mu}}u\cdot\partial\mu\right]\,.
\end{align}
To eliminate the temporal derivative, we employ the ideal hydrodynamic equation of motion. Using the thermodynamic relation $\varepsilon=Ts-p+\mu^*{n^*}$ and $n^*=\partial{p}/\partial{\mu^*}$, the conservation laws, i.e., $\partial_\mu{T^{\mu\nu}}=0$ and $\partial_\mu\left(n^*u^\mu\right)=0$, at $T=0$ give\footnote{Note in the relation $\partial_\mu\left(n^*u^\mu\right)=0$ we adopt $\mu^*$ for the Lagrange multiplier.}
\begin{align}
	\mu^*{u\cdot\partial}{u^\nu}+\Delta^{\mu\nu}\partial_\mu{\mu^*}=0\,.
\end{align}
Building upon this, the solution takes the form
\begin{align}
	\partial_\mu\ln\mu^*=u\cdot\partial{u_\mu}+\xi{u_\mu}\vartheta\,.
\end{align}
By applying the conservation law of the energy-momentum tensor, $u_\nu{\partial_\mu{T^{\mu\nu}}}=0$, we obtain
\begin{align}
	\xi=-\frac{n^*}{\dif{\varepsilon}/\dif{\mu^*}}\,.
\end{align}
Using the thermodynamic relation $n^*=\dif{p}/\dif{\mu^*}$ and the definition of the square of the sound velocity ${v^2_s}=\dif{p}/\dif{\varepsilon}$, we can show that
\begin{align}
	\xi=-\frac{\dif{p}/\dif{\mu^*}}{\dif{\varepsilon}/\dif{\mu^*}}=-{v^2_s}\,,
\end{align}
and therefore
\begin{align}
	\partial_\mu{\mu}=\mu^*\left(u\cdot\partial u_\mu-{v^2_s}u_\mu\vartheta\right)\frac{\dif{\mu}}{\dif{\mu^*}}\,.
\end{align}
From this expression, we can derive
\begin{align}
	\begin{split}
		\phi&=\frac{\tau_\text{rel}}{E^*_{\boldsymbol{p}}}\left[\left(p\cdot\partial\right)\left(u\cdot{p}\right)-\mu^*\left(u\cdot\partial\right)\left(u\cdot{p}\right)\right.\\
		&\quad\left.+{v^2_s}\mu^*\left(u\cdot{p}-m^*\frac{\dif{m^*}}{\dif{\mu^*}}\right)\vartheta\right]\,.
	\end{split}
\end{align}
As a result, the nonequilibrium correction $\delta f$ can be decomposed into
\begin{align}
	\delta{f}=\phi_1p^\rho{p^\sigma}\left(D_\rho{u_\sigma}+D_\sigma{u_\rho}+\frac{2}{3}\Delta_{\rho\sigma}\vartheta\right)-\phi_2\vartheta\,,
\end{align}
with coefficients
\begin{subequations}
    \begin{gather}
		\phi_1=\frac{\tau_\text{rel}}{2E^*_{\boldsymbol{p}}}\delta\left(\mu^*-u\cdot{p}\right)\,,\\
		\begin{split}
			\phi_2=\frac{2}{3}\phi_1\left\{\left(p\cdot{u}\right)^2\left[1-3v_s^2\left(1-\frac{m^*}{\mu^*}\frac{\dif{m^*}}{\dif{\mu^*}}\right)\right]-{m^*}^2\right\}\,.
		\end{split}
	\end{gather}
\end{subequations}
Note that in the nonrelativistic limit, the form of the solution can indeed be cast back to the ansatz given in Eq.~(\ref{eq:coadjoint-orbit}).
Considering that the viscous part of the energy-momentum tensor in the local rest frame of the fluid is given by
\begin{align}
	\Pi^{\mu\nu}=\int\frac{\dif^3\boldsymbol{p}}{\left(2\pi\right)^3}\frac{p^\mu{p^\nu}}{E^*_{\boldsymbol{p}}}\delta{f}\,,
\end{align}
we obtain expressions for the shear viscosity $\eta$ and bulk viscosity $\zeta$ as
\begin{subequations}
    \begin{gather}
		\eta=\frac{2}{15}\int\frac{\dif^3\boldsymbol{p}}{\left(2\pi\right)^3}\frac{\boldsymbol{p}^4}{E^*_{\boldsymbol{p}}}\phi_1\,,\\
		\zeta=\frac{1}{3}\int\frac{\dif^3\boldsymbol{p}}{\left(2\pi\right)^3}\frac{\boldsymbol{p}^2}{E^*_{\boldsymbol{p}}}\phi_2\,.
    \end{gather}
\end{subequations}

\subsection{Thermodynamics of quasiparticle system within mean field}\label{sec1b}
Before proceeding further, we would like to discuss some specific thermodynamic quantities of dense and cold fermions. For free fermions at zero temperature, the pressure $p_0$, energy density $\varepsilon_0$, and particle number density $n_0$ are given by~\cite{Li:2022url}
\begin{subequations}
	\label{eq:thermodynamic function}
	\begin{align}
		\begin{split}
			p_0(\mu,m) 
			&= \frac{2s+1}{48\pi^{2}} \left[ |\mu| \left(2\mu^2 - 5m^2\right) \sqrt{\mu^2 - m^2} \right. \\
			&\quad \left. + 3m^4 \cosh^{-1}\left(\frac{|\mu|}{m}\right) \right] \Theta(|\mu| - m)\,, \label{eq:thermodynamic-function1}
		\end{split} \\
		\begin{split}
			\varepsilon_0(\mu,m) 
			&= \frac{2s+1}{16\pi^2} \left[ |\mu| \left(2\mu^2 - m^2\right) \sqrt{\mu^2 - m^2} \right. \\
			&\quad \left. - m^4 \cosh^{-1}\left(\frac{|\mu|}{m}\right) \right] \Theta(|\mu| - m)\,, \label{eq:thermodynamic-function2}
		\end{split} \\
		n_0(\mu,m) 
		&= \left(2s+1\right) \frac{\left(\mu^2 - m^2\right)^{\frac{3}{2}}}{6\pi^2} \Theta\left(|\mu| - m\right)\,. \label{eq:thermodynamic-function3}
	\end{align}
\end{subequations}
with $s$ corresponding to the intrinsic spin of the fermion and the inverse hyperbolic cosine function expressed as $\cosh^{-1}\left(z\right)=\ln(z+\sqrt{z^2-1})$. Based on these expressions, the squared speed of sound $v_s^2$ as a function of $m$ and $\mu$ is calculated as\footnote{In Eqs. (\ref{eq:thermodynamic function}) and (\ref{eq:vs2}), the absolute value of the chemical potential is for the general expression.}
\begin{align}
	\label{eq:vs2}
	v_s^2=\frac{1}{3}\left(1-\frac{m^2}{\mu^2}\right)\Theta\left(\left|\mu\right|-m\right)\,.
\end{align}
As $m$ approaches zero, $v_s$ converges to the characteristic speed of sound in a massless Fermi gas, which is $1/\sqrt{3}$. With this, we have
\begin{align}
	\phi_2=\frac{\tau_\text{rel}}{3E^*_{\boldsymbol{p}}}\delta\left(\mu^*-p\cdot{u}\right)\left({\mu^*}^2-{m^*}^2\right)\frac{m^*}{\mu^*}\frac{\dif{m^*}}{\dif{\mu^*}}\,.
\end{align}
From this, we can notice that the sign of $\zeta$ is determined by $\dif{m^*}/\dif{\mu^*}$. In the MFA, the effective potential $\bar{\mathcal{V}}_\mathrm{eff}$ can generally be expressed as
\begin{align}
\begin{split}
	\bar{\mathcal{V}}_\mathrm{eff}\left(\mu;\bar{\sigma},\bar{\omega}_0\right)
	&=-p\left(\mu;\bar{\sigma},\bar{\omega}_0\right)\\
	&=-\sum_ip_0\left[m^*_i\left(\bar{\sigma}\right),\mu^*_i\left(\bar{\omega}\right)\right]+\mathcal{U}\left(\bar{\sigma},\bar{\omega}_0\right)\,,
\end{split}
\end{align}
where the subscript $i$ sums over all particle species. The residual interaction contributions in the MFA are encapsulated in the potential $\mathcal{U}$. The condensates $\bar{\sigma}$ and $\bar{\omega}_0$ are determined self-consistently by the gap equations obtained from extremum conditions
\begin{subequations}
	\label{eq:gap equation}
    \begin{gather}
    \mathcal{q}_1\left[m^*\left(\bar{\sigma}\right),\bar{\sigma},\mu^*\right]=\frac{\partial\bar{\mathcal{V}}_\mathrm{eff}}{\partial\bar{\sigma}}\equiv0\,,\label{eq:gap equation1}
		\\
\mathcal{q}_2\left[m^*,\mu^*\left(\bar{\omega}_0\right),\bar{\omega}_0\right]=\frac{\partial\bar{\mathcal{V}}_\mathrm{eff}}{\partial\bar{\omega}_0}\equiv0\,.\label{eq:gap equation2}
    \end{gather}
\end{subequations}
Furthermore, to ensure the stability of the thermodynamic system, the following conditions must be met:
\begin{subequations}
	\label{eq:stability_condition}
		\begin{gather}
			\begin{vmatrix}
				\displaystyle\frac{\partial^2 \bar{\mathcal{V}}_\mathrm{eff}}{\partial \bar{\sigma}^2} & \displaystyle\frac{\partial^2 \bar{\mathcal{V}}_\mathrm{eff}}{\partial \bar{\sigma} \partial \bar{\omega}_0} \\
				\displaystyle\frac{\partial^2 \bar{\mathcal{V}}_\mathrm{eff}}{\partial \bar{\omega}_0 \partial \bar{\sigma}} & \displaystyle\frac{\partial^2 \bar{\mathcal{V}}_\mathrm{eff}}{\partial \bar{\omega}_0^2}
			\end{vmatrix}>0\,,\label{eq:stability_condition1}
\\
			\frac{\partial^2 \bar{\mathcal{V}}_\mathrm{eff}}{\partial \bar{\sigma}^2}>0\,.\label{eq:stability_condition2}
		\end{gather}
\end{subequations}
To calculate $\dif m^* / \dif \mu^*$, we use the condition
\begin{align}
	\frac{\dif\mathcal{q}_1}{\dif{\mu}}=\frac{\dif\mathcal{q}_2}{\dif{\mu}}=0\,,
\end{align}
which leads to
\begin{subequations}
    \begin{gather}
		\left[\frac{\partial\mathcal{q}_1}{\partial{m^*}}+\frac{1}{\mathcal{f}^\prime\left(\bar{\sigma}\right)}\frac{\partial\mathcal{q}_1}{\partial\bar{\sigma}}\right]\frac{\dif{m^*}}{\dif\mu}+\frac{\partial\mathcal{q}_1}{\partial\mu^*}\frac{\dif{\mu^*}}{\dif\mu}=0\,,\\
		\frac{\partial\mathcal{q}_2}{\partial{m^*}}\frac{\dif{m^*}}{\dif\mu}+\left[\frac{\partial\mathcal{q}_2}{\partial{\mu^*}}+\frac{1}{\mathcal{g}^\prime\left(\bar{\omega}_0\right)}\frac{\partial\mathcal{q}_2}{\partial\bar{\omega}_0}\right]\frac{\dif{\mu^*}}{\dif\mu}=\frac{1}{\mathcal{g}^\prime\left(\bar{\omega}_0\right)}\frac{\partial\mathcal{q}_2}{\partial\bar{\omega}_0}\,.
    \end{gather}
\end{subequations}
Solving these equations, we find
\begin{subequations}
    \begin{gather}
		\frac{\dif{m^*}}{\dif\mu}=-\frac{\frac{\partial\mathcal{q}_1}{\partial\mu^*}\cdot \frac{1}{\mathcal{g}^\prime\left(\bar{\omega}_0\right)}\frac{\partial\mathcal{q}_2}{\partial\bar{\omega}_0}}{\left[\frac{\partial\mathcal{q}_1}{\partial{m^*}}+\frac{1}{\mathcal{f}^\prime\left(\bar{\sigma}\right)}\frac{\partial\mathcal{q}_1}{\partial\bar{\sigma}}\right] \left[\frac{\partial\mathcal{q}_2}{\partial{\mu^*}}+\frac{1}{\mathcal{g}^\prime\left(\bar{\omega}_0\right)}\frac{\partial\mathcal{q}_2}{\partial\bar{\omega}_0}\right]-\frac{\partial\mathcal{q}_1}{\partial\mu^*} \frac{\partial\mathcal{q}_2}{\partial{m^*}}}\,,\\
		\frac{\dif{\mu^*}}{\dif\mu}=\frac{\left[\frac{\partial\mathcal{q}_1}{\partial{m^*}}+\frac{1}{\mathcal{f}^\prime\left(\bar{\sigma}\right)}\frac{\partial\mathcal{q}_1}{\partial\bar{\sigma}}\right] \frac{1}{\mathcal{g}^\prime\left(\bar{\omega}_0\right)}\frac{\partial\mathcal{q}_2}{\partial\bar{\omega}_0}}{\left[\frac{\partial\mathcal{q}_1}{\partial{m^*}}+\frac{1}{\mathcal{f}^\prime\left(\bar{\sigma}\right)}\frac{\partial\mathcal{q}_1}{\partial\bar{\sigma}}\right] \left[\frac{\partial\mathcal{q}_2}{\partial{\mu^*}}+\frac{1}{\mathcal{g}^\prime\left(\bar{\omega}_0\right)}\frac{\partial\mathcal{q}_2}{\partial\bar{\omega}_0}\right]-\frac{\partial\mathcal{q}_1}{\partial\mu^*}\frac{\partial\mathcal{q}_2}{\partial{m^*}}}\,.
    \end{gather}
\end{subequations}
From this, we can solve for
\begin{align}
	\label{eq:dm*/dmu*}
	\frac{\dif{m^*}}{\dif{\mu^*}}=\frac{\dif{m^*}/\dif\mu}{\dif{\mu^*}/\dif\mu}=-\frac{\mathcal{f}^\prime\left(\bar{\sigma}\right)}{\mathcal{g}^\prime\left(\bar{\omega}_0\right)}\frac{\partial^2\bar{\mathcal{V}}_\mathrm{eff}}{\partial\bar{\omega}_0\partial\bar{\sigma}}\left(\frac{\partial^2\bar{\mathcal{V}}_\mathrm{eff}}{\partial\bar{\sigma}^2}\right)^{-1}\,.
\end{align}
With the quasi-Fermi momentum defined as $p^*_\mathrm{F}=\sqrt{{\mu^*}^2-{m^*}^2}$, the transport coefficients can be expressed as
\begin{subequations}\label{eq:expression_of_transport_coefficient}
		\begin{gather}
		\eta=\frac{{p^*_\mathrm{F}}^5}{30\pi^2\mu^*}\tau_\text{rel}\left(\mu^*\right)\,,\label{eq:expression_of_transport_coefficient1}\\
		\zeta=-\frac{{p^*_\mathrm{F}}^5}{18\pi^2{\mu^*}}\frac{m^*\mathcal{f}^\prime\left(\bar{\sigma}\right)}{\mu^*\mathcal{g}^\prime\left(\bar{\omega}_0\right)}\frac{\partial^2\bar{\mathcal{V}}_\mathrm{eff}}{\partial\bar{\omega}_0\partial\bar{\sigma}}\left(\frac{\partial^2\bar{\mathcal{V}}_\mathrm{eff}}{\partial\bar{\sigma}^2}\right)^{-1}\tau_\text{rel}\left(\mu^*\right)\,.\label{eq:expression_of_transport_coefficient2}
		\end{gather}
\end{subequations}
We note that the condition for the positive definiteness of the transport coefficients imposes that
\begin{align}\label{eq:dissipative_condition}
\frac{\mathcal{f}^\prime\left(\bar{\sigma}\right)}{\mathcal{g}^\prime\left(\bar{\omega}_0\right)}\frac{\partial^2\bar{\mathcal{V}}_\mathrm{eff}}{\partial\bar{\omega}_0\partial\bar{\sigma}}\left(\frac{\partial^2\bar{\mathcal{V}}_\mathrm{eff}}{\partial\bar{\sigma}^2}\right)^{-1} < 0
\end{align}
which is an independent constraint from the stability conditions of the thermal-dynamical potential given in Eqs.~(\ref{eq:stability_condition}). This is a somewhat surprising result. Because the positivity of transport coefficient ensures a positive definite local entropy production rate. Therefore, the evolution of the system is dissipative and stable. On the other hand, the stability of the global thermodynamic stability is given by Eqs.~(\ref{eq:stability_condition}). So we would imagining that the two sets of conditions are related to each other. However, Eq.~(\ref{eq:dissipative_condition}) and (\ref{eq:stability_condition}) are independent sets of conditions. It is possible that this is an artifact of the use of RTA of the collision term. In future works, we will go beyond the RTA collision terms and investigate the positivity condition of the bulk viscosity.

\subsection{Extraction of the relaxation time}\label{sec1c}
In this section, we extract the relaxation time $\tau_{\rm rel}$ used in the RTA collision term from the full collision term that includes $2 \leftrightarrow 2$ scattering processes of identical fermions. We designate the incoming particles as $a$ and $b$, and the outgoing particles as $c$ and $d$. In the previous formulas, $E_{\boldsymbol{p}}^*$ should now be replaced by $E_a^*$. For spin-$1/2$ fermions, the collision term can be expressed as
\begin{align}
	\label{eq:collision_term1}
	\begin{split}
		\mathcal{C}\left[f\right]=&\frac{1}{2}\frac{1}{2E_{a}^*}\int_{\boldsymbol{p}_b,\boldsymbol{p}_c,\boldsymbol{p}_d}\left[W\left(a+b\rightarrow{c+d}\right)f_cf_d\bar{f}_a\bar{f}_b\right.\\
		&\left.-W\left(c+d\rightarrow{a+b}\right)f_af_b\bar{f}_c\bar{f}_d\right]\,,
	\end{split}
\end{align}
where $\bar{f}_i=1-f_i$ and we have abbreviated the integration measure as $\int_{\boldsymbol{p}}\equiv\int\dif^3\boldsymbol{p}/\big[2E_{\boldsymbol{p}}\left(2\pi\right)^3\big]$, and $\int_{\boldsymbol{p}_1,\cdots,\boldsymbol{p}_n}\equiv\int\prod_{i=1}^{n}\dif^3\boldsymbol{p}_i/\big[2E_{\boldsymbol{p}_i}\left(2\pi\right)^3\big]$. The prefactor $1/2$ accounts for identical incoming particles. The collision kernel is given by
\begin{align}
	W\left(i+j\rightarrow{k+l}\right)=\frac{1}{4}\sum_{\rm spin}\left|\mathcal{M}\right|^2\left(2\pi\right)^4\delta\left(p_i+p_j-p_k-p_l\right)\,,
\end{align}
with $\mathcal{M}$ being the scattering amplitude. Considering unpolarized scattering, we sum over the spins in the initial and final states, and the factor $1/4$ is for averaging over initial spins. From the detailed balance principle, we have $W\left(a+b\leftrightarrow{c+d}\right)\equiv W\left(a+b\rightarrow{c+d}\right)=W\left(c+d\rightarrow{a+b}\right)$. To maintain consistency with the left-hand side of the Boltzmann equation, we perform a linear expansion of $\delta f_a$. Consequently, the collision term in Eq. (\ref{eq:collision_term1}) transforms into
\begin{align}\label{eq:Collision_Term}
	\mathcal{C}\left[f\right]=-\omega\left(E_a^*\right)\delta{f}_a\,,
\end{align}
where the typical relaxation frequency is calculated as\footnote{Henceforth, $f_{0i}$ will denote the equilibrium distribution function of particle $i$.}
\begin{align}
	\label{eq:relaxation_frequency}
	\begin{split}
		\omega\left(E_{a}^*\right)=&\frac{1}{2}\frac{1}{2E_{a}^*}\int_{\boldsymbol{p}_b,\boldsymbol{p}_c,\boldsymbol{p}_d}W\left(a+b\leftrightarrow{c+d}\right)\\
		&\times\left(f_{0c}f_{0d}\bar{f}_{0b}+f_{0b}\bar{f}_{0c}\bar{f}_{0d}\right)\,.
	\end{split}
\end{align}
From this, we can estimate the relaxation time via $\tau_\mathrm{rel}=1/\omega$.

Since all states are occupied under the Fermi surface in a fermionic system at exactly zero temperature, no state beyond this energy can be occupied. The collisions near the Fermi surface are forbidden by the conservation of energy,  resulting in a vanishing viscosity. The evolution of the system causes no increase in entropy, which is also compatible with the third law of thermodynamics of $s(T=0)=0$.

At a finite yet negligible temperature ($T/\mu^*\ll1$), the distribution functions can be expanded in powers of $T/\mu^*$ to determine $\omega$. This approach is analogous to applying the Sommerfeld expansion to an integral involving the Fermi--Dirac distribution~\cite{Sommerfeld:1928elektronentheorie}.
\begin{figure}[htpb!]
	\centering
	\includegraphics[width=0.5\textwidth]{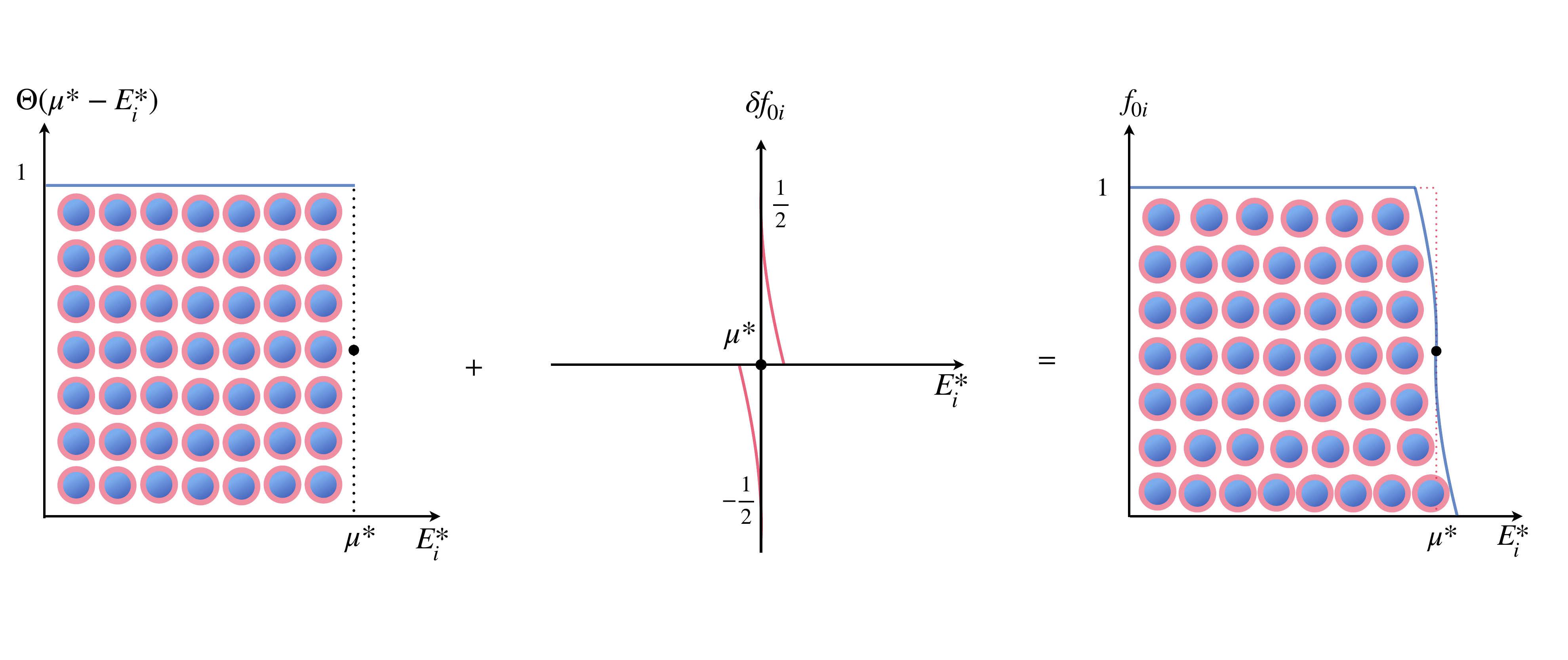}
	\caption{A sketch of the Sommerfeld expansion at $T/\mu^*\ll1$.
		\label{fig:Sommmerfeld_expansion}}
\end{figure}
Within this framework, the equilibrium distribution function can be decomposed:
\begin{align}\label{eq:fN1}
          f_{0i}&=\Theta\left(\mu^*-E_i^*\right)+\delta f_{0i}\,,
\end{align}
with
\begin{align}\label{eq:fN2}
\delta f_{0i}&=-\frac{\left(\pi T\right)^2}{6}\delta^\prime\left(E_i^*-\mu^*\right)+\mathcal{O}\left(\frac{T^4}{\mu^{*4}}\right)
\end{align} 
Figure~\ref{fig:Sommmerfeld_expansion} is a schematic plot of such an expansion.
In the product of particle distributions in Eq. (\ref{eq:relaxation_frequency}), the term consisting solely of the zeroth-order contributions vanishes due to the absence of available states to accommodate outgoing particles. Accordingly, the detailed balance process in the same order also vanishes. Beyond the zeroth order, it can be observed that the term at order $T^2$ is raised from $f_{0c}f_{0d}\bar{f}_{0b}$, as it primarily stems from the occupancy effect predominantly influenced by density. In contrast, the detailed balance term $f_{0b}\bar{f}_{0c}\bar{f}_{0d}$ contributes only from order $T^4$, since the inverse process is governed by thermal excitations. As a result, we retain only the leading-order contribution in the nonvanishing distributions, which is computed as
\begin{align}
	\label{eq:distribution_in_phase_space}
	\begin{split}
		f_{0c}f_{0d}\bar{f}_{0b}\approx&-\Theta\left(\mu^*-E_c^*\right)\Theta\left(\mu^*-E_d^*\right)\delta f_{0b}\\
		&+\Theta\left(\mu^*-E_d^*\right)\Theta\left(E_b^*-\mu^*\right)\delta{f_{0c}}\\
		&+\Theta\left(\mu^*-E_c^*\right)\Theta\left(E_b^*-\mu^*\right)\delta{f_{0d}}\,.
	\end{split}
\end{align}
The collision term is calculated in the rest frame of the fluid cell. More details of the calculation are provided in Appendix \ref{sec:Collision term}.
\section{Dense nucleon matter at low temperature}\label{sec2}
\subsection{Static properties}\label{sec2a}
Protons and neutrons serve as the primary fermionic degrees of freedom in this study. The nuclear force is inherently complex, featuring long-range meson-mediated attractions alongside short-range repulsive interactions that are pivotal for achieving the saturation density of nuclear matter. To accurately model these phenomena, we adopt the Walecka model \cite{Walecka:1974qa}, a relativistic mean-field theory renowned for its ability to describe the binding and saturation characteristics of nuclear systems. Through the inclusion of both scalar and vector meson exchanges, the Walecka model offers a comprehensive framework for investigating the equation of state of nuclear matter under diverse conditions \cite{Csernai:1986qf, Herbert:1992kd, DasGupta:2000md}.

The Lagrangian density of the Walecka model is composed of $\mathcal{L}_{\rm W}=\mathcal{L}_{\mathrm{N}}+\mathcal{L}_{\rm {m}}$ wherein the nucleon part $\mathcal{L}_{\mathrm{N}}$ and the meson part $\mathcal{L}_{\rm {m}}$ are, respectively, given by
\begin{subequations}\label{eq:Lagrangian density}
    \begin{gather}
        \mathcal{L}_{\mathrm{N}}=\sum_{\rm N=n,p} \bar{\psi}_{\rm N}\left(i \slashed{\partial}-m_{\rm N}+g_{\sigma } \sigma-g_{\omega } \slashed{\omega}+\mu_B\gamma_0\right) \psi_{\rm N}\,,\label{eq:Lagrangian density1}\\
		\mathcal{L}_{\rm {m}}=\frac{1}{2}\left(\partial_{\mu} \sigma \partial^{\mu} \sigma-m_{\sigma}^{2} \sigma^{2}\right)-U(\sigma)-\frac{1}{4} F^{\mu \nu} F_{\mu \nu}+\frac{1}{2} m_{\omega}^{2} \omega^{\mu} \omega_{\mu}\label{eq:Lagrangian density2}\,,
    \end{gather}
\end{subequations}
where $F_{\mu\nu}=\partial_\mu \omega_\nu-\partial_\nu \omega_\mu$ and $\psi_{\rm N}$ (${\rm N=n,p}$) denotes the nucleon fields, with $\mathrm{n}$ representing neutrons and $\mathrm{p}$ for protons. In the present frame, isospin symmetry is assumed for simplicity, so both of mass are $m_{\rm N}$. The scalar $\sigma$ meson with mass $m_\sigma$ and the vector $\omega$ meson with mass $m_\omega$ are included to account for the long-range attraction and short-range repulsion of the nuclear force, respectively. These interactions are captured by the meson-nucleon coupling terms, with coupling constants $g_\sigma$ and $g_\omega$. Additionally, we introduce a phenomenological potential~\cite{Kapusta:2006pm}
\begin{align}
	U(\sigma)=\frac{1}{3} b m_{\rm N}\left(g_{\sigma} \sigma\right)^{3}+\frac{1}{4} c\left(g_{\sigma} \sigma\right)^{4}
\end{align}
to fit the empirically observed properties of nuclear matter. Currently, the conserved charge is the net baryon number $B$, so we then specify the chemical potential and particle density as baryonic, i.e., $\mu\rightarrow\mu_B$, $\mu^*\rightarrow\mu^*_B$, $n^*\rightarrow n^*_B$ and the mass as that of the nucleon, i.e., $m^*\rightarrow m^*_\mathrm{N}$.

The partition function of the model can be expressed in the imaginary-time path integral formalism:
\begin{align}
	\begin{split}
		{\cal Z}_\mathrm{W}=&\int \prod_{\mathrm{N}=\mathrm{n},\mathrm{p}}[\dif\psi_\mathrm{N}][\dif\bar{\psi}_\mathrm{N}][\dif\sigma][\dif\omega_\mu]\\
		&\times e^{\int_0^{\beta} \text{d}\tau \int \text{d}^3{\boldsymbol{x}} \left({\cal L}_{\rm W}+\mu_B\sum_{\mathrm{N}=\mathrm{n},\mathrm{p}}\psi_\mathrm{N}^\dagger\psi_\mathrm{N}^{\phantom{\dag}}\right)}\,.
	\end{split}
\end{align}
Based on the relation
\begin{align}
	{\cal V}_\mathrm{W}\left(T,\mu_B;\bar\sigma,\bar\omega_0\right)=\frac{T}{V}\ln{\cal Z}_\mathrm{W}\,,
\end{align}
the effective potential of the Walecka model is given by
\begin{align}
	\begin{split}
		{\cal V}_\mathrm{W}\left(T,\mu_B\right)=&-\frac{T}{V}\sum_{n}\int\frac{\dif^3\boldsymbol{k}}{\left(2\pi\right)^3}\sum_{\mathrm{N}=\mathrm{n},\mathrm{p}}\\
		&\times\ln\det\left[{\cal S}_{0\mathrm{N}}^{-1}\left(ik_n,{\boldsymbol{k}})+\Sigma_\mathrm{N}(\sigma,\omega_0\right)\right]\\
		&+\frac{1}{2}m_\sigma^2\sigma^2+U\left(\sigma\right)-\frac{1}{2}m_\omega^2\omega^\mu\omega_\mu\,,
	\end{split}
\end{align}
where ${\cal S}_{\rm0\mathrm{N}}^{-1}=\left(ik_n+\mu_B\right)\gamma_0-\boldsymbol{\gamma}\cdot\boldsymbol{k}-m_\mathrm{N}$ is the inverse of the thermal Green's function of free nucleons, with $k_n=(2n+1)\pi T$ ($n \in \mathbb{Z}$). The quantities $\Sigma_\mathrm{N}$ are defined as $\Sigma_{\rm n}=\Sigma_{\rm p}=g_{\sigma }\sigma-g_{\omega }\slashed{\omega}$.

Because of the symmetry spontaneous breaking, the meson field can be decomposed into condensate and fluctuations, i.e., $\sigma=\bar{\sigma}+\delta\sigma$ and $\omega^\mu=\bar{\omega}_0+\delta\omega^\mu$. In the MFA, the fluctuations $\delta\sigma$ and $\delta\omega^\mu$ are neglected. At zero temperature, the equation of state in MFA of the Walecka model can be evaluated as
\begin{align}
	\begin{split}
		\bar{p}_\mathrm{W}\left(\mu_B\right)&=-\bar{{\cal V}}_\mathrm{W}\left(\mu_B\right)\\
		&=2p_0\left(\mu^*_B, m^*_\mathrm{N}\right)-\frac{1}{2} m_{\sigma}^{2} \bar{\sigma}^{2}-U(\bar{\sigma}) +\frac{1}{2} m_{\omega}^{2} \bar{\omega}_{0}^{2}\,,
	\end{split}
\end{align}
where the effective masses of quasinucleon $m^*_\mathrm{N}$ and quasi-Fermi surface $\mu^*_B$ are
\begin{subequations}
	\begin{gather}
	m^*_\mathrm{N}=m_{\rm N}-g_{\sigma}\bar{\sigma}\,,\\
	\mu^*_B=\mu_B-g_{\omega} \bar{\omega}_{0}\,.
    \end{gather}
\end{subequations}
Note that the radiative correction from the vacuum contribution has been neglected. The meson condensates $\bar{\sigma}$ and $\bar{\omega}_0$ are determined by the following gap equations based on Eqs.~(\ref{eq:gap equation}):
\begin{subequations}\label{eq:gap_equations}
	\begin{gather}
		-m_{\omega}^{2} \bar{\omega}_{0}+2 g_{\omega} n_0\left(\mu^*_B, m^*_\mathrm{N}\right)=0\,,\\
		m_{\sigma}^{2} \bar{\sigma}+U^\prime(\bar{\sigma})-2 g_{\sigma} n_{\rm s}\left(\mu^*_B, m^*_\mathrm{N}\right)=0\,.
	\end{gather}
\end{subequations}
Here, $n_{\rm s}(\mu,m)$ is for the free scalar density defined as
\begin{align}
	\begin{split}
		n_{\rm s}\left(\mu,m\right)=&\frac{m}{2\pi^2}\left[\left|\mu\right|\sqrt{\mu^2-m^2}-m^2\cosh^{-1}\left(\frac{\left|\mu\right|}{m}\right)\right]\\
		&\times\Theta\left(\left|\mu\right|-m\right)\,.
	\end{split}
\end{align}
The gap equations presented here account for the self-energy corrections to both the nucleon mass and the Fermi surface, resulting from interactions with bare mesons. In more self-consistent approaches, one ought to incorporate medium effects on the mesons, such as vacuum polarization. A more comprehensive treatment, including the two-particle irreducible (2PI) formalism through the Cornwall–Jackiw–Tomboulis approach, has been investigated in Refs. \cite{Shu:2005cw,Shu:2007vr}. While the 2PI framework provides a more complete and self-consistent description, the MFA effectively captures the primary static thermodynamic properties of nucleons at low temperature and moderate baryon chemical potential. Additionally, the medium-induced corrections to $m_\sigma$ and $m_\omega$ at low temperature are relatively small compared to their vacuum masses \cite{Shu:2007vr}. Since this study does not aim to provide a precise depiction of in-medium nucleon behavior, we have chosen to neglect meson collective oscillations in medium and other many-body effects in our calculations.

\begin{figure}[htbp!]
	\centering
	\includegraphics[width=0.48\textwidth]{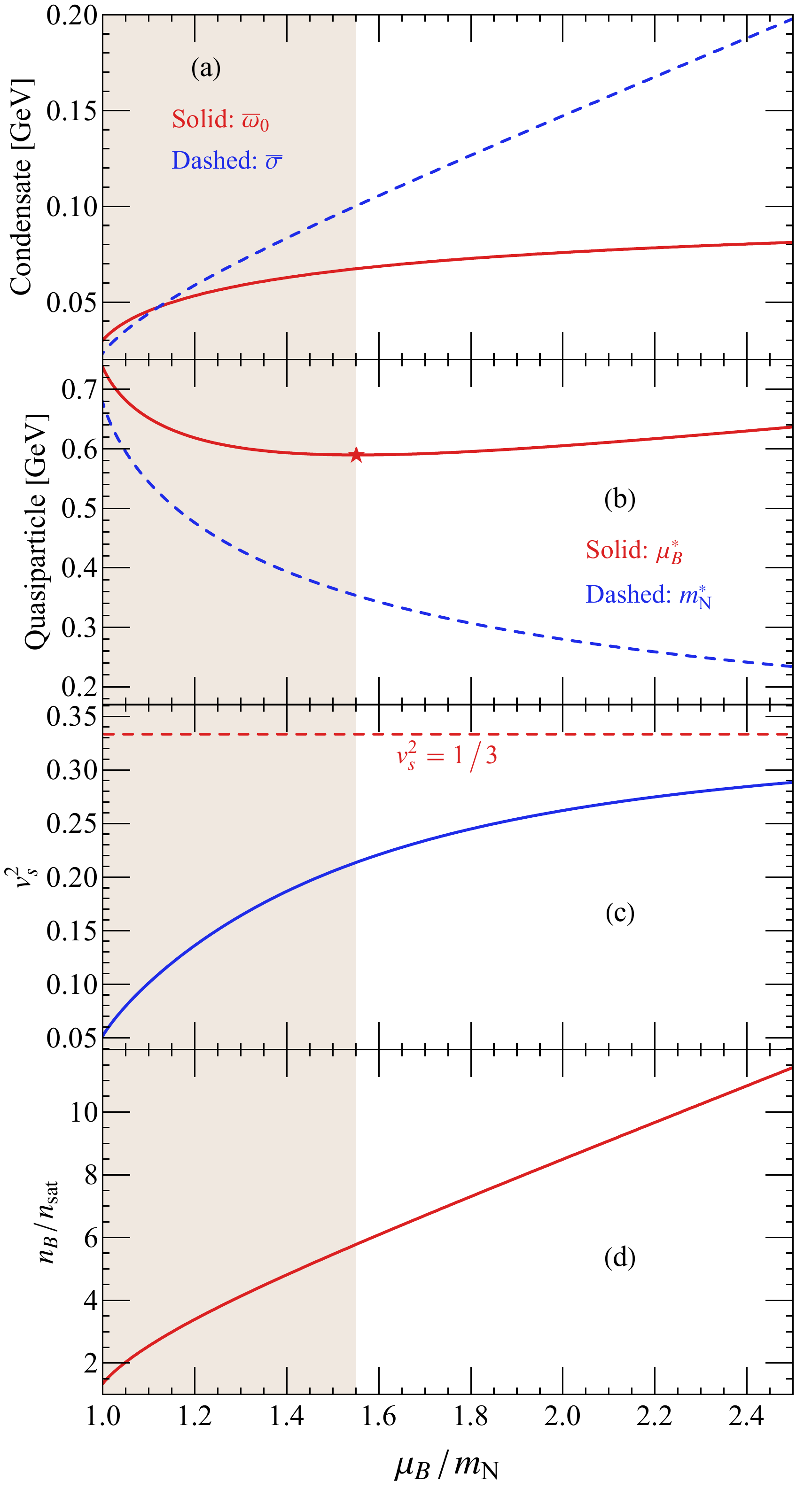}
	\caption{
		Static properties of nucleon matter depicted by the Walecka model presented as functions of $\mu_B/m_\mathrm{N}$: (a) $\bar{\omega}_0$ (red, solid) and $\bar{\sigma}$ (blue, dashed); (b) $\mu_B^*$ (red, solid) and $m^*_\mathrm{N}$ (blue, dashed), where the red star indicates the inflection point of $\mu_B^*$; (c) squared sound velocity $v_s^2$ (blue, solid) compared to the benchmark $v_s^2 = 1/3$ (red, dashed); (d) Baryon number density $n_B = 2n_0\left(\mu_B^*,m_\mathrm{N}^*\right)$, scaled by the nuclear saturation density $n_\mathrm{sat} \approx 0.16~\text{fm}^{-3}$. The brown shaded area represents the physical region predicted by the Walecka model based on Eq. (\ref{eq:Waleck_dmdmuB}).
		\label{fig:static}}
\end{figure}

In our calculations, we set the model parameters as follows: The particle masses are taken to be  $m_{\rm N}=0.939$ GeV, $m_\sigma=0.550$ GeV, and $m_\omega=0.783$ GeV~\cite{ParticleDataGroup:2022pth}. The coupling constants are chosen as $g_\sigma^2/(4\pi)=6.003$, $g_\omega^2/(4\pi)=5.948$, $b=7.950\times10^{-3}$, and $c=6.952\times10^{-4}$ to fit the empirically known properties of nuclear matter~\cite{Kapusta:2006pm}. To find all solutions of the gap equations, we apply the homotopy method to solve this nonlinear equation group~\cite{chen2015homotopy, Wang:2012me}. The static properties of nucleons at $T = 0$ are shown in Fig.~\ref{fig:static}.
In Fig. \ref{fig:static}(b), as the density increases, an inflection point in $\mu_B^*$ appears, i.e., $\dif{\mu_B^*}/\dif{\mu_B}=0$, marked by a red star. This will lead to a divergence at $\mu_B\approx1.551m_\mathrm{N}$ in the derivative
\begin{widetext}
	\begin{align}
		\label{eq:Waleck_dmdmuB}
		\frac{\dif{m^*_\mathrm{N}}}{\dif{\mu_B^*}}=-\frac{2g_\sigma^2m^*_\mathrm{N}p^*_\mathrm{F}}{g_\sigma^2\mu^*_Bp^*_\mathrm{F}+\pi^2\left[m_\sigma^2+U^{\prime\prime}\left(\bar{\sigma}\right)\right]-3g_\sigma^2{m^*_\mathrm{N}}^2\cosh^{-1}\left(\mu^*_B/m^*_\mathrm{N}\right)}\,,
	\end{align}
\end{widetext}
which is calculated from Eq.~(\ref{eq:dm*/dmu*}). This divergence occurs because, although the Walecka model in the MFA can characterize the gas-liquid phase transition of nucleon matter, the ground state is not stable. The elements of the Hessian matrix are
\begin{subequations}\label{eq:element_Hession_matrix}
\begin{gather}
		\frac{\partial^2\bar{\mathcal{V}}_\mathrm{W}}{\partial\bar{\sigma}^2}=U^{\prime\prime}\left(\bar{\sigma}\right)+\frac{g_\sigma^2 \left[\mu^*_Bp_\mathrm{F}^*-3 m^{*2}_\mathrm{N} \cosh ^{-1}\left(\frac{\mu^*_B}{m^*_\mathrm{N}}\right)\right]}{\pi ^2}+m_\sigma^2\,,\label{eq:element_Hession_matrix1}\\
		\frac{\partial^2\bar{\mathcal{V}}_\mathrm{W}}{\partial\bar{\omega}^2}=-\frac{2 g_\omega^2 \mu^*_Bp_\mathrm{F}^*}{\pi ^2}-m_\omega^2\,,\label{eq:element_Hession_matrix2}\\
		\frac{\partial^2\bar{\mathcal{V}}_\mathrm{W}}{\partial\bar{\omega}\partial\bar{\sigma}}=\frac{\partial^2\bar{\mathcal{V}}_\mathrm{W}}{\partial\bar{\sigma}\partial\bar{\omega}}=\frac{2g_\sigma g_\omega m^*_\mathrm{N} p_\mathrm{F}^*}{\pi^2}\,.\label{eq:element_Hession_matrix3}
\end{gather}
\end{subequations}
Apparently, $\partial^2\bar{\mathcal{V}}_\mathrm{W}/\partial\bar{\omega}^2$ in Eq. (\ref{eq:element_Hession_matrix2}) is always negative, while the off-diagonal elements in Eq. (\ref{eq:element_Hession_matrix3}) are positive. In Eq. (\ref{eq:element_Hession_matrix1}), we have
\begin{align}
	\lim\limits_{{\mu_B^*}/{m_\mathrm{N}^*}\rightarrow+\infty}\frac{\partial^2\bar{\mathcal{V}}_\mathrm{W}}{\partial\bar{\sigma}^2}>0\,,
\end{align}
yet the Hessian matrix in Eq. (\ref{eq:stability_condition1}) cannot be guaranteed to be positive-definite. Therefore, even though the gap equations are solved, the stability condition specified by Eqs.~(\ref{eq:stability_condition}) is not satisfied when ${\mu_N^*}/{m_\mathrm{N}^*}\gg1$. This suggests that the Walecka model, though widely used for the study of equilibrium properties of the nuclear matter, may not be suitable for the study of the dynamics properties, where the system can be driven out of local equilibrium. As in the discussion of Eqs.~(\ref{eq:element_Hession_matrix}), the physical region (brown shaded area in Fig. \ref{fig:static}) depicted by the Walecka model is defined as the region where Eq. (\ref{eq:Waleck_dmdmuB}) is positive. According to Fig. \ref{fig:static}(d), the Walecka model remains stable for $n_B \lesssim 6n_\mathrm{sat}$.
\subsection{Transport coefficients}\label{sec2b}
Since the contribution of the one-pion exchange potential to the bulk properties of nuclear matter largely averages to zero~\cite{Walecka:1974qa, Miller:1972zza}, we consider only the exchange of the $\sigma$ and $\omega$ mesons within the framework of the Walecka model. In the MFA, the system is treated as a free fermion gas, with part of the nonperturbative interaction effects incorporated into the effective nucleon mass $m^*_\mathrm{N}$ and effective baryon chemical potential $\mu^*_B$. In this context, the degree of freedom is associated with the quasinucleon field ${\psi}_\mathrm{N}^*$. However, in such an MFA picture, scattering processes are absent. Therefore, we need to go beyond the MFA.

We introduce the meson fluctuation fields $\delta\sigma$ and $\delta\omega^\mu$ by decomposing the meson fields around their condensate as
\begin{subequations}
	\begin{gather}
		\sigma=\bar{\sigma}+\left(\sigma-\bar{\sigma}\right)=\bar{\sigma}+\delta\sigma\,,\\
		\omega^\mu=\bar{\omega}_0+\left(\omega^\mu-\bar{\omega}_0\delta^{\mu0}\right)=\bar{\omega}_0+\delta\omega^\mu\,.
	\end{gather}
\end{subequations}
With $\psi_\mathrm{N}^*$, the Lagrangian density in Eq.~(\ref{eq:Lagrangian density}) can be decomposed as $\mathcal{L}_\mathrm{W}=\mathcal{L}_\mathrm{N}^*+U\left(\bar{\sigma}\right)+\delta\mathcal{L}_{\mathrm{N-m}}^*+\delta\mathcal{L}_\mathrm{m}$, where the fermionic component is given by
\begin{align}
	\mathcal{L}_\mathrm{N}^*=\sum_{\rm N=n,p} \bar{\psi}_{\rm N}^*\left(i \slashed{\partial}-m^*_\mathrm{N}+\mu_B^*\gamma^0\right) \psi_{\rm N}^*\,,
\end{align}
and the nucleon-meson interactions beyond the MFA are contained in the term
\begin{align}
	\delta\mathcal{L}_{\mathrm{N-m}}^*=\sum_{\rm N=n,p}\bar{\psi}_\mathrm{N}^*\left(g_\sigma\delta\sigma-g_\omega\delta\slashed{\omega}\right){\psi}_\mathrm{N}^*\,.
\end{align}
The meson fluctuation related terms are included in $\delta\mathcal{L}_\mathrm{m}$.

Based on Eqs. (\ref{eq:fN1}) and (\ref{eq:fN2}), we decompose $\psi_\mathrm{N}^*$ into two distinct components: core nucleons below the Fermi surface, $\psi_{\mathrm{N,c}}^*\equiv\psi_{\mathrm{N}}^*[E^*\leq\mu_B^*]$, which contribute to the thermodynamics of the ground state within the MFA, and nucleons near the Fermi surface, $\psi_{\mathrm{N,s}}^*\equiv\psi_{\mathrm{N}}^*[E^*\approx\mu_B^*]$, which participate in scattering processes that drive dissipation. As discussed in Sec. \ref{sec1c}, both energy and particle number conservation prohibits the scattering between $\psi_{\mathrm{N,c}}^*$ and $\psi_{\mathrm{N,s}}^*$. And considering that scattering processes are localized near the Fermi surface, $\mu_B^*$ for $\psi_\mathrm{N,s}^*$ can be absorbed by redefining the zero-energy level and streamlining the treatment of quasiparticle excitations. The quasinucleon Lagrangian density can be cleanly decomposed into two parts ${\mathcal{L}}_\mathrm{N}^*=\bar{\mathcal{L}}_\mathrm{N}^*+\delta{\mathcal{L}}_\mathrm{N}^*$, where the MFA component is
\begin{align}
	\bar{\mathcal{L}}_\mathrm{N}^*=\sum_{\rm N=n,p} \bar{\psi}_{\mathrm{N,c}}^*\left(i \slashed{\partial}-m^*_\mathrm{N}+\mu_B^*\gamma^0\right) \psi_{\mathrm{N,c}}^*\,,
\end{align}
and the scattering contribution is
\begin{align}
	\begin{split}
		\delta{\mathcal{L}}_\mathrm{N}^*=\sum_{\rm N=n,p} \bar{\psi}_{\mathrm{N,s}}^*\left(i \slashed{\partial}-m^*_\mathrm{N}\right) \psi_{\mathrm{N,s}}^*\,.
	\end{split}
\end{align}
Together with $U\left(\bar{\sigma}\right)$, $\bar{\mathcal{L}}_\mathrm{N}^*$ fully characterizes the thermodynamic properties of the ground state in the MFA. Then, $\delta\mathcal{L}_{\mathrm{N-m}}^*$ can also be decomposed into core and Fermi-surface contributions as
\begin{align}
	\delta\mathcal{L}_{\mathrm{N-m}}^*=\sum_{\mathrm{N=n,p}\atop i=\mathrm{c,s}}\bar{\psi}_{\mathrm{N},i}^*\left(g_\sigma\delta\sigma-g_\omega\delta\slashed{\omega}\right){\psi}_{\mathrm{N},i}^*\,,
\end{align}
To properly handle the  mesonic linear fluctuation terms, we examine the gap equations in the MFA, as presented in Eqs.~(\ref{eq:gap_equations}). These yield the following expressions\footnote{$\langle \cdots \rangle$ denotes the statistical average over the thermodynamic ensemble.}:
\begin{subequations}\label{eq:counter_term}
\begin{gather}
	m_{\omega}^{2} \bar{\omega}_{0}\delta\omega_0=2 g_{\omega} \langle{{\psi}_{\mathrm{N,c}}^*}^\dagger\psi_{\mathrm{N,c}}^*\rangle\delta\omega_0\,,\\
	m_{\sigma}^{2} \bar{\sigma}\delta\sigma+U^\prime(\bar{\sigma})\delta\sigma=2 g_{\sigma}\langle\bar{\psi}_{\mathrm{N,c}}^*\psi_{\mathrm{N,c}}^*\rangle\delta\sigma\,.
\end{gather}
\end{subequations}
Here we have used the definitions $n_0\equiv\langle{{\psi}_{\mathrm{N,c}}^*}^\dagger\psi_{\mathrm{N,c}}^*\rangle$ and $n_\mathrm{s}\equiv\langle\bar{\psi}_{\mathrm{N,c}}^*\psi_{\mathrm{N,c}}^*\rangle$ in MFA. Crucially, from Eqs.~(\ref{eq:counter_term}), we learn that the linear terms of meson fluctuations in $\delta\mathcal{L}_\mathrm{m}$ introduce the counterterms required to cancel the self-energy contributions in forward scattering for $\psi_{\mathrm{N,c}}^*$, and then only $\psi_{\mathrm{N,s}}^*$ contribute to the scattering with meson fluctuations in $\delta\mathcal{L}_{\mathrm{N-m}}^*$.

Neglecting meson fluctuations beyond the quadratic terms, the mesonic fluctuation Lagrangian density simplifies to
\begin{align}
	\begin{split}
		\delta{\mathcal{L}}_\mathrm{m}=\frac{1}{2}\left(\partial^\mu\delta\sigma\partial_\mu\delta\sigma-\tilde{m}_\sigma^2\delta\sigma^2\right)-\frac{1}{4}\delta{F}^{\mu\nu}\delta{F}_{\mu\nu}+\frac{1}{2}m_\omega^2\delta\omega^\mu\delta\omega_\mu\,,
	\end{split}
\end{align}
where $\delta{F}^{\mu\nu}=\partial^\mu\delta\omega^\nu+\partial^\nu\delta\omega^\mu$. The effective mass of the $\sigma$ meson fluctuation is modified by the potential term into
\begin{align}
	\tilde{m}_\sigma=\sqrt{m_\sigma^2+2bg_\sigma^3m_\mathrm{N}\bar{\sigma}+3cg_\sigma^4\bar{\sigma}^2}\,.
\end{align}
Therefore, the mesonic fluctuation fields $\delta\sigma$ and $\delta\omega^\mu$ inherit the same dynamical behavior as prescribed by the original mesonic Lagrangian density.

\begin{figure}[htpb!]
	\centering
	\includegraphics[width=0.47\textwidth]{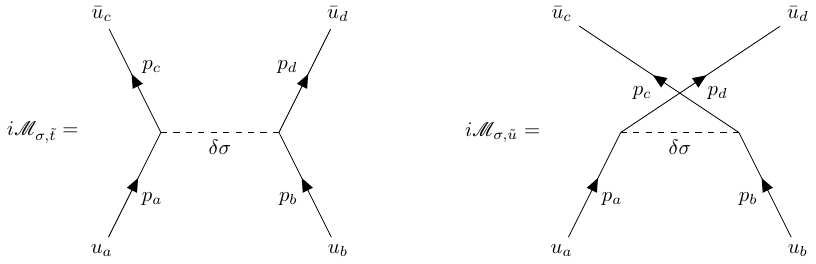}\\~\\
	\includegraphics[width=0.47\textwidth]{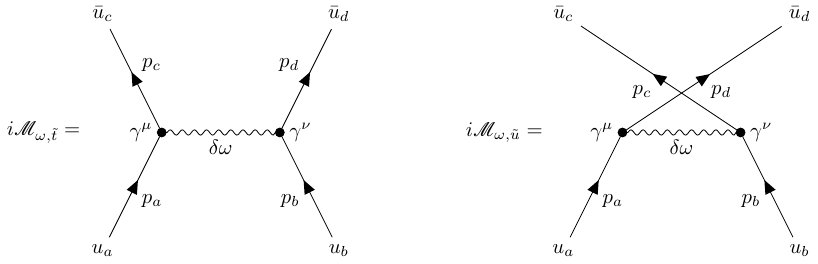}
	\caption{		
		Feynman diagrams of leading-order nucleon-nucleon scattering including the exchange of the fluctuations of $\sigma$ meson (upper panel) and the $\omega$ meson (lower panel). Only the $t$ channel (left panel) and $u$ channel (right panel) contribute. The incoming and outgoing nucleons with four-momentum $p_i$ ($i=a,b,c,d$) are denoted by spinors $u_i=u\left(p_i\right)$ and $\bar{u}_i=\bar{u}\left(p_i\right)$, respectively.
        \label{fig:nucleon_scattering}
	}
\end{figure}

The scattering processes that drive dissipation can be interpreted as quasifree nucleons near the Fermi surface interacting with meson fluctuations. These fluctuations effectively mimic the original meson field dynamics, with the key distinction being the modified mass $\tilde{m}_\sigma$ in the scalar sector. In this paper, we consider only the leading-order scattering, i.e., $2\leftrightarrow2$ tree-level processes \cite{Baym:1975va}. Therefore, only the $t$ and $u$ channels are present, as illustrated in Fig. \ref{fig:nucleon_scattering}.\footnote{To avoid repeating labels, we represent the Mandelstam variables using $\tilde{\mathcal{o}}$, where $\mathcal{o} = s, t, u$.}
The total scattering amplitude is given by
\begin{align}
	i\mathcal{M}=i\left[\left(\mathcal{M}_{\sigma,\tilde{t}}-\mathcal{M}_{\sigma,\tilde{u}}\right)+\left(\mathcal{M}_{\omega,\tilde{t}}-\mathcal{M}_{\omega,\tilde{u}}\right)\right]\,,
\end{align}
where the elements of the squared scattering amplitude are presented in Appendix \ref{sec:Elements of scattering amplitude}.

\begin{figure}[htpb!]
	\centering
	\includegraphics[width=0.47\textwidth]{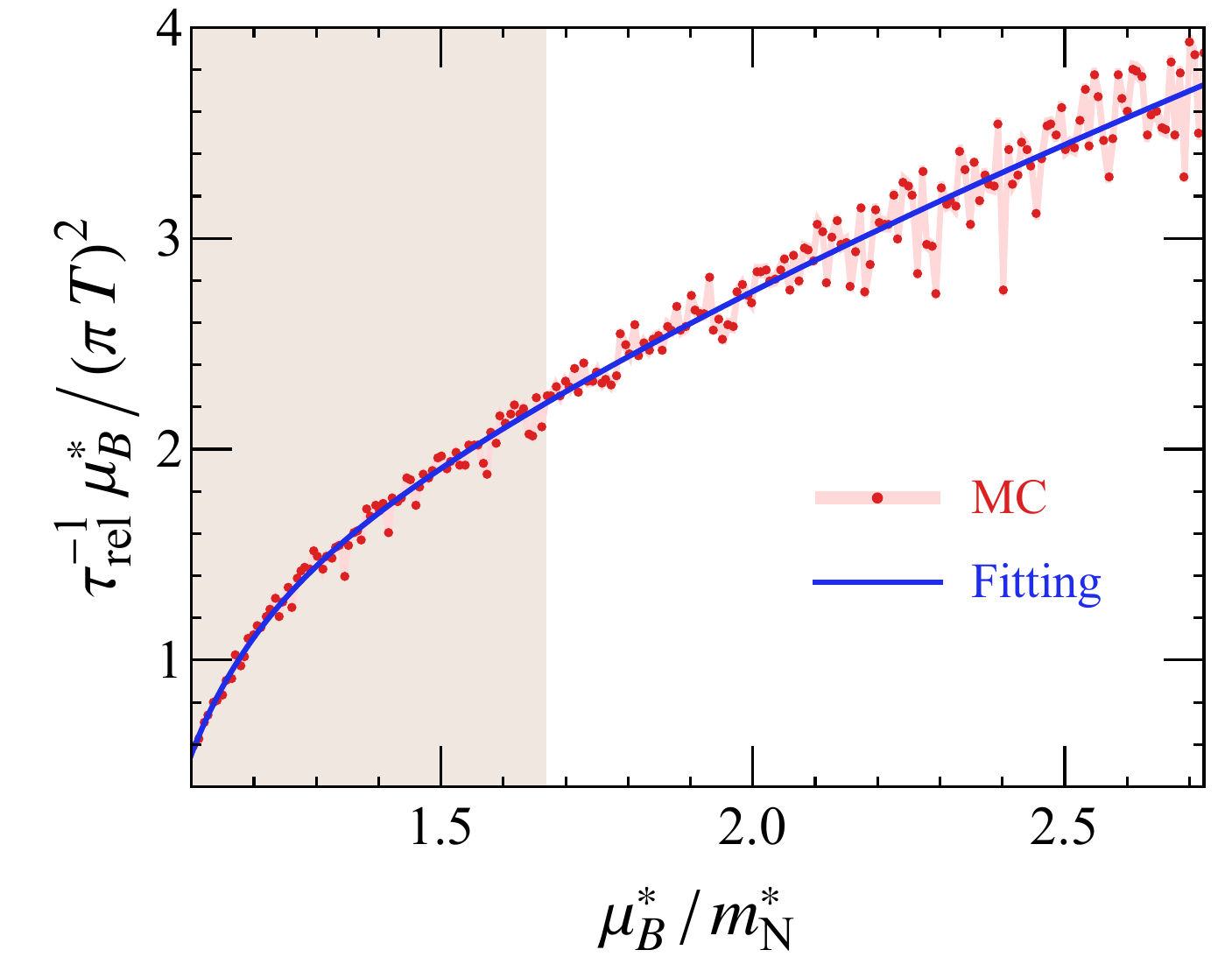}
	\caption{The scaled relaxation time $\tau_\mathrm{rel}^{-1} \mu_B^*/(\pi T)^2$ as a function of $\mu_B^*/m^*_\mathrm{N}$. The red band is calculated from Monte Carlo (MC) integration~\cite{Lepage:2020tgj}. The blue solid line is a fit to the MC data given in Eq.~(\ref{eq:fit_tau}). The brown area also represents the physical region calculated based on the results shown in Fig. \ref{fig:static}.
		\label{fig:relaxation_time}}
\end{figure}

With this scattering amplitude, we can estimate the relaxation time $\tau_\mathrm{rel}$. Considering that the binding energy of nucleons in heavy nuclei is around 8--10 MeV~\cite{Krane:1991introductory}, we set the temperature to $T=8$ MeV to avoid thermal excitation. The scaled relaxation time is shown in Fig.~\ref{fig:relaxation_time},
where the relaxation time is fitted as
\begin{align}
	\label{eq:fit_tau}
    \tau_\mathrm{rel}\left(x\right)=1966.8\frac{x^4-2.37x^3+5.44 x^2-5.63 x+1.90}{x^5+2.25 x^4-3.60 x^2}\,,
\end{align}
with $x=\mu^*_B/m^*_\mathrm{N}>1$. In this physical region, the relaxation time of dense and cold nucleon matter is of the order $\mu^*_B/\left(\pi T\right)^2$. At low density, i.e., small $\mu_B$, the relaxation time is large because the collision integral over phase space becomes zero if nucleon-bound states are not formed. Under this condition, the relaxation time diverges to infinity. As the density increases, the mean-free path reduces gradually, resulting in more frequent collisions within a given time period; therefore, the relaxation time decreases.

\begin{figure}[htpb!]
	\centering
	\includegraphics[width=0.47\textwidth]{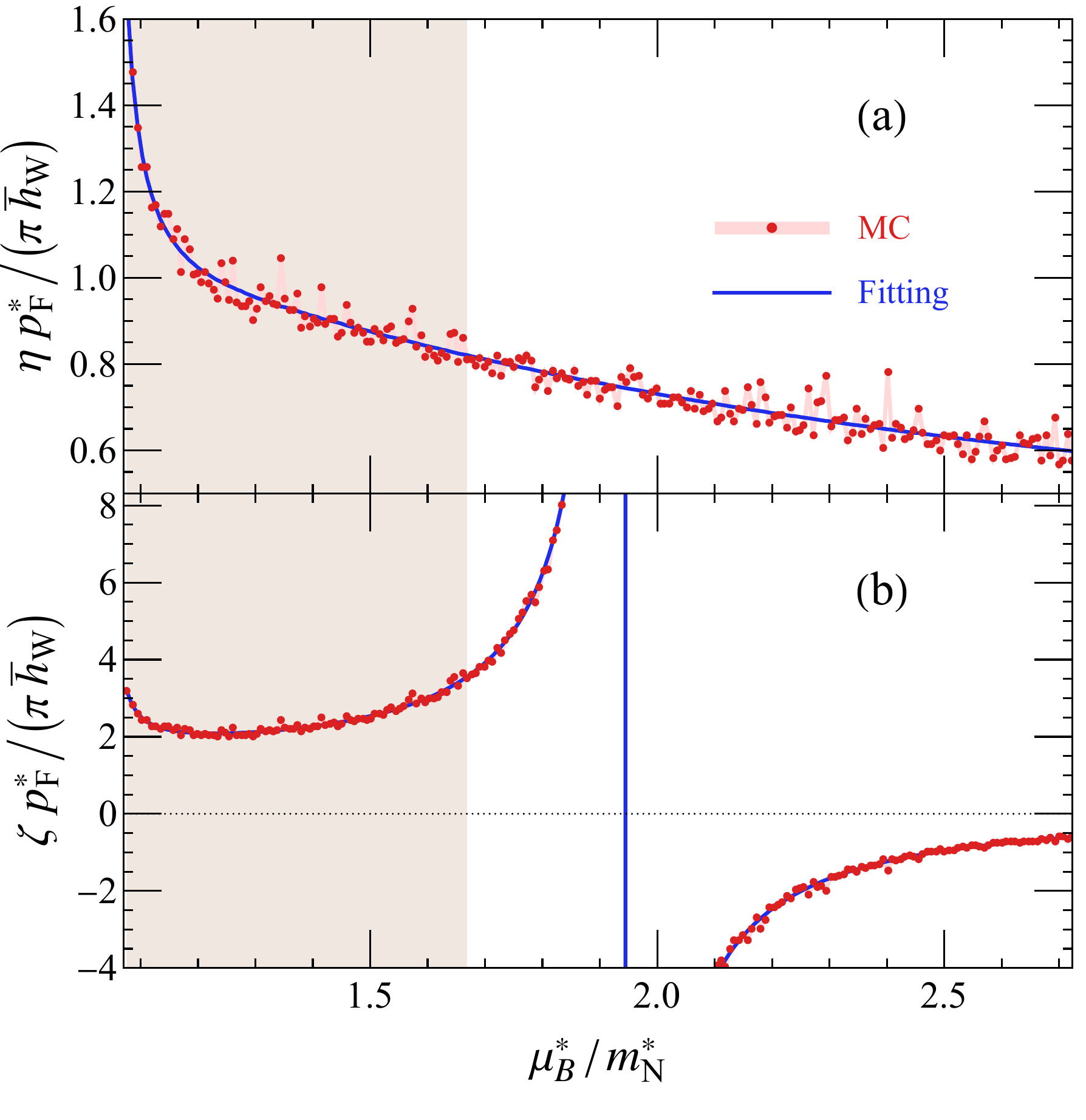}
	\caption{The dimensionless scaled (a) shear viscosity $\eta p_\mathrm{F}^*/\left(\pi\bar{h}_\mathrm{W}\right)$ and (b) bulk viscosity $\zeta p_\mathrm{F}^*/\left(\pi\bar{h}_\mathrm{W}\right)$ as functions of $\mu_B^*/m^*_\mathrm{N}$. The labels are the same to Fig. \ref{fig:relaxation_time}.
	\label{fig:transport_coefficients}}
\end{figure}

At finite baryon density, the fluidity of the system should be assessed based on the ratio of the transport coefficients to the enthalpy~\cite{Liao:2009gb, Du:2024wjm}. In the MFA at $T = 0$,\footnote{We consider the condition $T/\mu_B^*\ll1$, where both the static properties and the thermodynamic quantities at $T = 0$ serve as a good approximation at leading order expansion of $f_{0}$.} the energy density is evaluated as
\begin{align}
	\bar{\varepsilon}_\mathrm{W}\left(\mu^*_B,m^*_\mathrm{N}\right)=2\varepsilon_0\left(\mu^*_B,m^*_\mathrm{N}\right)+\frac{1}{2} m_{\sigma}^{2} \bar{\sigma}^{2}+U(\bar{\sigma}) +\frac{1}{2} m_{\omega}^{2} \bar{\omega}_{0}^{2}\,,
\end{align}
and the enthalpy density is calculated as
\begin{align}
	\bar{h}_\mathrm{W}\left(\mu^*_B,m^*_\mathrm{N}\right)=\bar{p}_\mathrm{W}\left(\mu^*_B,m^*_\mathrm{N}\right)+\bar{\varepsilon}_\mathrm{W}\left(\mu^*_B,m^*_\mathrm{N}\right)\,.
\end{align}
The dimensionless enthalpy-scaled shear viscosity and bulk viscosity are shown in Fig. \ref{fig:transport_coefficients}.\footnote{In fact, $\zeta$ will diverge rapidly with a sign change near the boundary of the brown area, which is not apparent in the current figure due to the limitations of the numerical calculation step size and the displayed range.}
In Fig. \ref{fig:transport_coefficients}(a), the scaled shear viscosity always decreases with increasing density, even beyond the physical region. The shear viscosity measures the resistance of fluid to shear flow. At higher densities, interactions among particles may facilitate easier sliding past one another because of the decreasing mean-free path, reducing the shear viscosity. In contrast, the bulk viscosity in Figure \ref{fig:transport_coefficients}(b) is nonmonotonic even within the physical region. This is because the bulk viscosity depends not only on microscopic interactions but also on complex static thermodynamic properties reflected in the derivative terms in Eq.~(\ref{eq:expression_of_transport_coefficient2}). Both the divergence and the negative values of the bulk viscosity in the nonphysical region result from the instability of the Walecka model, as mentioned in Eqs.~(\ref{eq:element_Hession_matrix}). We also notice that the bulk viscosity is nearly twice the shear viscosity. This suggests that in the cold and dense nucleon matter, processes involving compression or expansion are more heavily damped than those involving shear, indicating stronger interactions or relaxation processes that specifically resist volumetric deformation. The mechanisms for dissipating energy are more effective when dealing with changes in volume rather than shape.
\section{Conclusion and outlook}\label{sec3}
In this paper, we employed the Boltzmann equation framework within the Fermi liquid picture to investigate the transport properties of dense matter at extremely low temperatures. At leading order in the gradient expansion of the particle distribution function, we observed that fluctuations are primarily concentrated near the Fermi surface. Utilizing the RTA, we derived general expressions for both shear and bulk viscosities.

The static properties of nucleon matter were described using the Walecka model in the mean-field approximation. Our findings indicate that, at leading order in scattering, the relaxation time scales as $\tau_\mathrm{rel} \sim \mu_B^*/(\pi T)^2$. The fluidity of the cold and dense nucleon matter is characterized through shear and bulk viscosities of the order $p_\mathrm{F}^*/(\pi \bar{h}_\mathrm{W})$. Notably, the bulk viscosity is approximately twice the shear viscosity, suggesting that dissipation in the system more readily occurs through volumetric deformation rather than shear processes.

During the formulation of the transport coefficients, we observed that an isotropic Fermi surface at zero temperature prohibits collisions because all states below the Fermi energy are occupied, leaving no available states for fermions to scatter into above the Fermi surface. To address this, we introduced a negligible temperature to slightly deform the Fermi surface, enabling collisions. Fortunately, such deformation naturally occurs in nucleon matter due to spin polarization~\cite{Bertsch1978:collision, DABROWSKI1976452} and in neutron stars where magnetic fields affect the nucleons~\cite{Frick:2002fs}, where the transport coefficients at exact zero temperature can be studied.

Moreover, as we have pointed out, the Walecka model possesses inherent instabilities. To accurately describe nucleons and their interactions, more detailed calculations based on nucleon-nucleon scattering are necessary~\cite{Kaplan:1996xu,Wen:2023oju}. Due to the energy dependence of the relaxation time, it is essential to evaluate the bulk viscosity within the novel relaxation time approximation when solving the linearized Boltzmann equation in future work, ensuring compatibility with both micro- and macroscopic conservation laws \cite{Rocha:2021zcw, Hu:2024tnn}. Additionally, since the nuclei system is of finite size, the clear separation of length scales required for hydrodynamics may be absent. To study energy dissipation in such systems, it is essential to investigate response regimes beyond hydrodynamics~\cite{Ke:2022tqf,Brewer:2022ifw,Ke:2023bei, Gavassino:2023odx}.

Beyond nuclear physics, due to the artificially controlled scattering length in Feshbach resonance experiments, the transport coefficients in finite-size systems can also be examined in cold fermion atom systems~\cite{Floerchinger:2021ygn,Brandstetter:2023jsy}. These avenues offer promising directions for future research into the transport properties of dense fermionic systems.
\section{Acknowledgment}
The authors are grateful to Lipei Du, Defu Hou, Jin Hu, Fabian Rennecke, Andrey Sadofyev, Shuzhe Shi, and Yi Yin for their valuable discussions and comments. We also thank Jin Hu and Niko Jokela for pointing out the useful references. J. L. is supported by Grant No. 2004HBBHJD054.
\appendix
\begin{widetext}
\section{Collision term}\label{sec:Collision term}
In this appendix, we present the detailed calculation of the collision term. Based on Eq.~(\ref{eq:distribution_in_phase_space}), we separate $\omega\left(E_a^*\right)$ in Eq. (\ref{eq:relaxation_frequency}) into
\begin{align}
	\omega\left(E_a^*\right)=-\omega_{cd\delta{b}}+\omega_{d\tilde{b}\delta{c}}+\omega_{c\tilde{b}\delta{d}}\,.
\end{align}
Throughout this calculation, we set $\boldsymbol{p}_a$ along the $z$ direction, i.e.,
\begin{align}
	p_a^\mu=\left(E_a^*,0,0,\left|\boldsymbol{p}_a\right|\right)\,.
\end{align}

In the local rest frame of the fluid cell, the first term contributing to $\omega\left(E_a^*\right)$ is calculated as\footnote{The hat symbol ($\hat{\mathcal{o}}$) indicates that the variable ($\mathcal{o}$) should be expressed as a function of other independent variables.}
\begin{align}
	\omega_{cd\delta{b}}=\frac{\left(\pi T\right)^2}{24p_\mathrm{F}^*}\frac{\dif}{\dif\left|\boldsymbol{p}_b\right|}\left.\mathcal{F}_1\left(\left|\boldsymbol{p}_a\right|,\left|\hat{\boldsymbol{p}}_c\right|,\Omega_b,\Omega_c\right)\right|_{\left|\boldsymbol{p}_b\right|=p_\mathrm{F}^*\atop E_a^*=\mu^*}\,,
\end{align}
where
\begin{align}\label{F1}
	\mathcal{F}_1\left(\left|\boldsymbol{p}_a\right|,\left|\hat{\boldsymbol{p}}_c\right|,\Omega_b,\Omega_c\right)=\iint\dif\Omega_b\dif\Omega_c\frac{\left|\boldsymbol{p}_b\right|\left|\hat{\boldsymbol{p}}_c\right|}{32\hat{E}_d^*\left(2\pi\right)^5}\left|\mathcal{M}\right|^2\Theta\left(\mu^*-\hat{E}_c^*\right)\Theta\left(\mu^*-\hat{E}_d^*\right)\,.
\end{align}
Here, we introduce a new notation:
\begin{align}
	\hat{E}_i^*=\sqrt{\hat{\boldsymbol{p}}_i^2+{m^*}^2}\,.
\end{align}
Generally, in this section, $\Omega_i$ represents the solid angle between the $i$th momentum to the $z$ direction. The other two four-momenta are defined as
\begin{subequations}
	\begin{gather}
		p_b^\mu=\left(E_b^*,\left|\boldsymbol{p}_b\right|\sin\theta_b\cos\phi_b,\left|\boldsymbol{p}_b\right|\sin\theta_b\sin\phi_b,\left|\boldsymbol{p}_b\right|\cos\theta_b\right)\,,\\
		\hat{p}_c^\mu=\left(\hat{E}_c^*,\left|\hat{\boldsymbol{p}}_c\right|\sin\theta_c\cos\phi_c,\left|\hat{\boldsymbol{p}}_c\right|\sin\theta_c\sin\phi_c,\left|\hat{\boldsymbol{p}}_c\right|\cos\theta_c\right)\,.
	\end{gather}
\end{subequations}
By solving the equation
\begin{align}\label{eq:collision_eq1}
	\tilde{s}+2\left|\hat{\boldsymbol{p}}_c\right|\left|\boldsymbol{p}_{a+b}\right|\cos\hat{\theta}_{a+b}^c=2E^*_{a+b}\sqrt{\hat{\boldsymbol{p}}_c^2+{m^*}^2}\,,
\end{align}
we obtain\footnote{The redundant solution for $\left|\hat{\boldsymbol{p}}_c\right|$ is discarded based on the condition $\left|\hat{\boldsymbol{p}}_c\right|>0$.}
\begin{align}
		\left|\hat{\boldsymbol{p}}_c\right|=\frac{\left|\boldsymbol{p}_{a+b}\right|{\tilde{s}}\cos \hat{\theta}_{a+b}^c+E_{a+b}^*\sqrt{2 {m^*}^2 \left({\boldsymbol{p}_{a+b}^2}-2E_{a+b}^{*2}\right)+2 {m^*}^2 {\boldsymbol{p}_{a+b}^2} \cos2\hat{\theta}_{a+b}^c+{\tilde{s}}^2}}{2 \left({E_{a+b}^{*2}}-{\boldsymbol{p}_{a+b}^2} \cos ^2\hat{\theta}_{a+b}^c\right)}\,,
\end{align}
with
\begin{subequations}
	\begin{gather}
		\tilde{s}=\left(p_a+p_b\right)^2\,,\\
		E_{a+b}^*=E_a^*+E_b^*\,,\\
		\left|\boldsymbol{p}_{a+b}\right|=\left|\boldsymbol{p}_a+\boldsymbol{p}_b\right|=\sqrt{\boldsymbol{p}_a^2+\boldsymbol{p}_b^2+2\left|\boldsymbol{p}_a\right|\left|\boldsymbol{p}_b\right|\cos\theta_b}\,,\\
		\hat{\theta}_{a+b}^c=\cos^{-1}\left[\frac{\cos\theta_c\left(\left|\boldsymbol{p}_b\right|\cos\theta_b+\left|\boldsymbol{p}_a\right|\right)+\left|\boldsymbol{p}_b\right|\sin \theta_b \sin \theta_c \cos \left(\phi_b-\phi_c\right)}{\left|\boldsymbol{p}_{a+b}\right|}\right]\,.
	\end{gather}
\end{subequations}
To compute Eq.~(\ref{F1}), we use the relation
\begin{align}
	\left|\hat{\boldsymbol{p}}_d\right|=\sqrt{{E_{a+b}^*}\left(E_{a+b}^*-2\hat{E}_c^*\right)+\hat{\boldsymbol{p}}_c^2}\,.
\end{align}

Here, $\theta_i$ and $\phi_i$ are the polar and azimuthal angles, respectively, and $\hat{\theta}_i^j$ represents the polar angle between $i$th and the $j$th momentum. Although the subscripts may change in the following calculations, their physical meanings remain the same, and we will not clarify them again.

Similarly, the second part can be calculated as
\begin{align}
	\omega_{d\tilde{b}\delta{c}}=\frac{\left(\pi T\right)^2}{24p_\mathrm{F}^*}\frac{\dif}{\dif\left|\boldsymbol{p}_c\right|}\left.\mathcal{F}_2\left(\left|\boldsymbol{p}_a\right|,\left|\boldsymbol{p}_c\right|,\Omega_b,\Omega_c\right)\right|_{\left|\boldsymbol{p}_c\right|=p_\mathrm{F}^*\atop E_a^*=\mu^*}\,,
\end{align}
with
\begin{align}
	\mathcal{F}_2\left(\left|\boldsymbol{p}_a\right|,\left|\boldsymbol{p}_c\right|,\Omega_b,\Omega_c\right)=\iint\dif\Omega_b\dif\Omega_c\frac{\left|\hat{\boldsymbol{p}}_b\right|\left|\boldsymbol{p}_c\right|}{32\hat{E}_d^*\left(2\pi\right)^5}\left|\mathcal{M}\right|^2\Theta\left(\mu^*-\hat{E}_d^*\right)\Theta\left(\hat{E}_b^*-\mu^*\right)\,.
\end{align}
With
\begin{subequations}
	\begin{gather}
		\hat{p}_b^\mu=\left(\hat{E}_b^*,\left|\hat{\boldsymbol{p}}_b\right|\sin\theta_b\cos\phi_b,\left|\hat{\boldsymbol{p}}_b\right|\sin\theta_b\sin\phi_b,\left|\hat{\boldsymbol{p}}_b\right|\cos\theta_b\right)\,,\\
		p_c^\mu=\left(E_c^*,\left|\boldsymbol{p}_c\right|\sin\theta_c\cos\phi_c,\left|\boldsymbol{p}_c\right|\sin\theta_c\sin\phi_c,\left|\boldsymbol{p}_c\right|\cos\theta_c\right)\,,
	\end{gather}
\end{subequations}
and by solving the equation
\begin{align}\label{eq:collision_eq2}
	\tilde{t}+2\left|\hat{\boldsymbol{p}}_b\right|\left|\boldsymbol{p}_{c-a}\right|\cos\hat{\theta}_{c-a}^b=2E^*_{c-a}\sqrt{\boldsymbol{p}_b^2+{m^*}^2}\,,
\end{align}
other variables are expressed as
\begin{align}
	\left|\hat{\boldsymbol{p}}_b\right|=\frac{\left|\boldsymbol{p}_{c-a}\right| \tilde{t} \cos\hat{\theta}_{c-a}^b-E_{c-a}^* \sqrt{2{m^*}^2 \left(\boldsymbol{p}_{c-a}^2-2E_{c-a}^{*2}\right)+2{m^*}^2 \boldsymbol{p}_{c-a}^2 \cos2\hat{\theta}_{c-a}^b+\tilde{t}^2}}{2 \left(E_{c-a}^{*2}-\boldsymbol{p}_{c-a}^2 \cos ^2\hat{\theta}_{c-a}^b\right)}\,,
\end{align}
and
\begin{subequations}
	\begin{gather}
		\tilde{t}=\left(p_c-p_a\right)^2\,,\\
		E^*_{c-a}=E^*_c-E^*_a\,,\\
		\left|\boldsymbol{p}_{c-a}\right|=\left|\boldsymbol{p}_c-\boldsymbol{p}_a\right|=\sqrt{\boldsymbol{p}_a^2+\boldsymbol{p}_c^2-2 \left|\boldsymbol{p}_a\right| \left|\boldsymbol{p}_c\right| \cos \theta_c}\\
		\hat{\theta}_{c-a}^b=\cos^{-1}\left[\frac{\cos \theta_b \left(\left|\boldsymbol{p}_c\right| \cos \theta_c-\left|\boldsymbol{p}_a\right|\right)+\left|\boldsymbol{p}_c\right| \sin \theta_b \sin \theta_c \cos \left(\phi_b-\phi_c\right)}{\left|\boldsymbol{p}_{c-a}\right|}\right]\,.
	\end{gather}
\end{subequations}
Moreover, we have
\begin{align}
	\left|\hat{\boldsymbol{p}}_d\right|=\sqrt{E^*_{c-a}\left(E^*_{c-a}-2 \hat{E}^*_b\right)+\hat{\boldsymbol{p}}_b^2}\,.
\end{align}

The third part is calculated as
\begin{align}
	\omega_{c\tilde{b}\delta{d}}=\frac{\left(\pi T\right)^2}{24p_\mathrm{F}^*}\frac{\dif}{\dif\left|\boldsymbol{p}_d\right|}\left.\mathcal{F}_3\left(\left|\boldsymbol{p}_a\right|,\left|\boldsymbol{p}_d\right|,\Omega_b,\Omega_d\right)\right|_{\left|\boldsymbol{p}_d
		\right|=p_\mathrm{F}^*\atop E_a^*=\mu^*}\,,
\end{align}
with
\begin{align}
	\mathcal{F}_3\left(\left|\boldsymbol{p}_a\right|,\left|\boldsymbol{p}_d\right|,\Omega_b,\Omega_d\right)=\iint\dif\Omega_b\dif\Omega_d\frac{\left|\hat{\boldsymbol{p}}_b\right|\left|\boldsymbol{p}_d\right|}{32\hat{E}_c^*\left(2\pi\right)^5}\left|\mathcal{M}\right|^2\Theta\left(\mu^*-\hat{E}_c^*\right)\Theta\left(\hat{E}_b^*-\mu^*\right)\,.
\end{align}
With the independent momenta written as
\begin{subequations}
	\begin{gather}
		\hat{p}_b^\mu=\left(\hat{E}_b^*,\left|\hat{\boldsymbol{p}}_b\right|\sin\theta_b\cos\phi_b,\left|\hat{\boldsymbol{p}}_b\right|\sin\theta_b\sin\phi_b,\left|\hat{\boldsymbol{p}}_b\right|\cos\theta_b\right)\,,\\
		p_d^\mu=\left(E_d^*,\left|\boldsymbol{p}_d\right|\sin\theta_d\cos\phi_d,\left|\boldsymbol{p}_d\right|\sin\theta_d\sin\phi_d,\left|\boldsymbol{p}_d\right|\cos\theta_d\right)\,,
	\end{gather}
\end{subequations}
and by solving the equation
\begin{align}
	\label{eq:collision_eq3}
	\tilde{u}+2\left|\hat{\boldsymbol{p}}_b\right|\left|\boldsymbol{p}_{d-a}\right|\cos\hat{\theta}_{d-a}^b=2E^*_{d-a}\sqrt{\boldsymbol{p}_b^2+{m^*}^2}\,,
\end{align}
other variables are expressed as
\begin{subequations}
	\begin{gather}
        \left|\hat{\boldsymbol{p}}_b\right|=\frac{\left|\boldsymbol{p}_{d-a}\right| \tilde{u} \cos\hat{\theta}_{d-a}^b-E_{d-a}^* \sqrt{2{m^*}^2 \left(\boldsymbol{p}_{d-a}^2-2E_{d-a}^{*2}\right)+2{m^*}^2 \boldsymbol{p}_{d-a}^2 \cos2\hat{\theta}_{d-a}^b+\tilde{u}^2}}{2 \left(E_{d-a}^{*2}-\boldsymbol{p}_{d-a}^2 \cos ^2\hat{\theta}_{d-a}^b\right)}\,,\\
		\tilde{u}=\left(p_d-p_a\right)^2\,,
		\\
		E^*_{d-a}=E^*_d-E^*_a\,,\\
		\left|\boldsymbol{p}_{d-a}\right|=\left|\boldsymbol{p}_d-\boldsymbol{p}_a\right|=\sqrt{\hat{\boldsymbol{p}}_a^2+\hat{\boldsymbol{p}}_d^2-2\left|\hat{\boldsymbol{p}}_a\right|\left|\hat{\boldsymbol{p}}_d\right|\cos\theta_d}\\
		\hat{\theta}_{d-a}^b=\cos^{-1}\left[\frac{\cos \theta_b \left(\left|\boldsymbol{p}_d\right| \cos \theta_d-\left|\boldsymbol{p}_a\right|\right)+\left|\boldsymbol{p}_d\right| \sin \theta_b \sin \theta_d \cos \left(\phi_d-\phi_b\right)}{\left|\boldsymbol{p}_{d-a}\right|}\right]\,.
	\end{gather}
\end{subequations}
and also
\begin{align}
	\left|\hat{\boldsymbol{p}}_c\right|=\sqrt{E^*_{d-a}\left(E^*_{d-a}-2 \hat{E}^*_b\right)+\hat{\boldsymbol{p}}_b^2}\,.
\end{align}

\section{Calculation of the squared scattering amplitude}
\label{sec:Elements of scattering amplitude}
Using \textsc{FeynCalc} 9.3.1~\cite{Mertig:1990an,Shtabovenko:2016sxi,Shtabovenko:2020gxv}, the elements in $\sum_{\rm spins}\left|\mathcal{M}\right|^2$ are calculated as
\begin{subequations}
    \begin{gather}
	\sum_{\rm spins}\left|\mathcal{M}_{\sigma,\tilde{t}}\right|^2=\frac{4g_\sigma^4}{\left(\tilde{t}-\tilde{m}_\sigma^2\right)^2}\left(\tdt-4{m^*_\mathrm{N}}^2\right)^2\,,\\	
	\sum_{\rm spins}\left|\mathcal{M}_{\sigma,\tilde{u}}\right|^2=\frac{4g_\sigma^4}{\left(\tilde{u}-\tilde{m}_\sigma^2\right)^2}\left(\tdu-4{m^*_\mathrm{N}}^2\right)^2\,,\\
	\sum_{\rm spins}\left|\mathcal{M}_{\omega,\tilde{t}}\right|^2=\frac{4g_\omega^4}{\left(\tilde{t}-m_\omega^2\right)^2}\left[\left(\tds+\tdt\right)^2+\left(\tdt+\tdu\right)^2-\left(\tds+\tdu\right)^2+\left(\tds^2+\tdt^2+\tdu^2\right)\right]\,,\\
	\sum_{\rm spins}\left|\mathcal{M}_{\omega,\tilde{u}}\right|^2=\frac{4g_\omega^4}{\left(\tilde{u}-m_\omega^2\right)^2}\left[\left(\tds+\tdu\right)^2+\left(\tdt+\tdu\right)^2-\left(\tds+\tdt\right)^2+\left(\tds^2+\tdt^2+\tdu^2\right)\right]\,,\\
	\sum_{\rm spins}\Re\left(\mathcal{M}_{\sigma,\tilde{t}}\mathcal{M}_{\sigma,\tilde{u}}^*\right)=\frac{g_\sigma^4}{\left(\tilde{t}-\tilde{m}_\sigma^2\right)\left(\tilde{u}-\tilde{m}_\sigma^2\right)}\left[\left(\tds+\tdt\right)^2+\left(\tds+\tdu\right)^2-\left(\tdt+\tdu\right)^2\right]\,,\\
	\sum_{\rm spins}\Re\left(\mathcal{M}_{\omega,\tilde{t}}\mathcal{M}_{\omega,\tilde{u}}^*\right)=\frac{2g_\omega^4}{\left(\tilde{t}-m_\omega^2\right)\left(\tilde{u}-m_\omega^2\right)}\left[\left(\tds+\tdt\right)^2+\left(\tds+\tdu\right)^2-3\left(\tdt+\tdu\right)^2-\left(\tds^2+\tdt^2+\tdu^2\right)\right]\,,\\
	\sum_{\rm spins}\Re\left(\mathcal{M}_{\sigma,\tilde{t}}\mathcal{M}_{\omega,\tilde{t}}^*\right)=-\frac{16g_\sigma^2g_\omega^2}{\left(\tilde{t}-\tilde{m}_\sigma^2\right)\left(\tilde{t}-m_\omega^2\right)}{m^*_\mathrm{N}}^2\left(\tds-\tdu\right)\,,\\
	\sum_{\rm spins}\Re\left(\mathcal{M}_{\sigma,\tilde{u}}\mathcal{M}_{\omega,\tilde{u}}^*\right)=-\frac{16g_\sigma^2g_\omega^2}{\left(\tilde{u}-\tilde{m}_\sigma^2\right)\left(\tilde{u}-m_\omega^2\right)}{m^*_\mathrm{N}}^2\left(\tds-\tdt\right)\,,\\
	\sum_{\rm spins}\Re\left(\mathcal{M}_{\sigma,\tilde{t}}\mathcal{M}_{\omega,\tilde{u}}^*\right)=-\frac{g_\sigma^2g_\omega^2}{\left(\tilde{t}-\tilde{m}_\sigma^2\right)\left(\tilde{u}-m_\omega^2\right)}
	\left[4\left(\tds+\tdu\right)^2+\left(\tdt+\tdu\right)^2-\left(\tds+\tdt\right)^2+\tdu^2-\tds^2\right]\,,\\
	\sum_{\rm spins}\Re\left(\mathcal{M}_{\sigma,\tilde{u}}\mathcal{M}_{\omega,\tilde{t}}^*\right)=-\frac{g_\sigma^2g_\omega^2}{\left(\tilde{u}-\tilde{m}_\sigma^2\right)\left(\tilde{t}-m_\omega^2\right)}\left[4\left(\tds+\tdt\right)^2+\left(\tdt+\tdu\right)^2-\left(\tds+\tdu\right)^2+\tdt^2-\tds^2\right]\,.
    \end{gather}
\end{subequations}
\end{widetext}
\bibliography{Ref}
\end{document}